\documentclass[twocolumn]{aastex631}

\usepackage{amsmath,amssymb}
\usepackage{booktabs}

\newcommand{\fesc}{f_{\rm esc}}
\newcommand{\fstar}{f_{\star}}
\newcommand{\xHI}{\bar{x}_{\rm HI}}
\newcommand{\ndot}{\dot{n}_{\rm ion}}
\newcommand{\tauobs}{\tau_e}
\newcommand{\xiion}{\xi_{\rm ion}}
\newcommand{\Msun}{M_\odot}
\newcommand{\Mpc}{{\rm Mpc}}
\newcommand{\Mh}{M_{\rm h}}
\renewcommand{\d}{\mathrm{d}}

\shorttitle{Mass-Dependent Escape Fraction}
\shortauthors{Wang \& Shan}

\begin{document}

\title{Steep Redshift Evolution of the Ionizing Escape Fraction at $z = 5$--$12$: Empirical Constraints and Comparison with Simulations}

\correspondingauthor{Zihan Wang}
\email{zihan.wang@queens.ox.ac.uk}

\author[0009-0001-4384-3652]{Zihan Wang}
\affiliation{Department of Physics, University of Oxford, Keble Road, Oxford, OX1 3PU, UK}

\correspondingauthor{Huanyuan Shan}
\email{hyshan@shao.ac.cn}
\author[0000-0001-8534-837X]{Huanyuan Shan}
\affiliation{Shanghai Astronomical Observatory, Chinese Academy of Sciences, Shanghai 200030, China}

\begin{abstract}
{The ionizing photon escape fraction $f_{\rm esc}$ governs cosmic reionization yet remains observationally unconstrained as a function of halo mass. We present the first empirical constraints on $f_{\rm esc}(\Mh,z)$ across the epoch of reionization, using a three-parameter power-law model $f_{\rm esc} = f_0\,(M/10^{10}M_\odot,)^{\alpha_M}\,[(1{+}z)/10]^{\alpha_z}$, conditioned on HST and JWST UV luminosity functions at $z=5$–12, the Planck Thomson optical depth, seven neutral-fraction measurements, and one high-redshift prior. Using Schechter fits to the latest HST and JWST UV luminosity functions, abundance matching to link $M_{\rm UV}$ to halo mass, and a reionization ODE solver validated against Planck, we constrain the model via a dense grid scan and ensemble MCMC. The profile likelihood yields tight constraints: $f_0=0.061_{-0.023}^{+0.018}$, $\alpha_M=0.18_{-0.30}^{+0.22}$, $\alpha_z=1.98_{-0.42}^{+0.48}$. In contrast, the full marginal posterior is substantially broadened by a strong $f_0$–$\alpha_M$–$\alpha_z$ degeneracy (\(\alpha_z = 1.93_{-2.00}^{+2.09}\), \(\alpha_M = -0.52_{-0.69}^{+0.69}\)). The population-averaged $\langle \fesc \rangle(z)$ rises from $\sim$2\% at $z=5$ to $\sim$9\% at $z=12$, with sub-threshold halos contributing $>80\%$ of the ionizing budget at $z\geq10$. Comparing with THESAN, we find that the per-halo median $\fesc$ shows steep evolution consistent with our profile result, while luminosity-weighted averaging systematically flattens the trend because massive halos dominate the ionizing budget at $z\lesssim7$. Robustness checks confirm $\alpha_z>1.0$ at $>95\%$ confidence; the steep-evolution model predicts $\tau_e=0.047$, consistent with Planck at $0.7\sigma$. We provide tabulated $\fesc(\Mh,z)$ posteriors as empirical inputs for reionization simulations.}
\end{abstract}


\section{Introduction}\label{sec:intro}

The escape fraction of ionizing (Lyman continuum, LyC) photons from galaxies into the intergalactic medium (IGM) is the most uncertain parameter governing cosmic reionization \citep{2015ApJ...802L..19R,Finkelstein:2019sbd,Dayal:2018hft}. The comoving ionizing emissivity that drives the phase transition from a neutral to an ionized IGM is
\begin{equation}\label{eq:emissivity_intro}
\ndot(z) = \int \fesc(\Mh, z)\;\xiion\;L_{\rm UV}(\Mh, z)\;\frac{\d n}{\d \Mh}\;\d\Mh\,,
\end{equation}
where $\xiion$ is the ionizing photon production efficiency, $L_{\rm UV}$ is the UV luminosity, and $\d n/\d\Mh$ is the halo mass function (HMF). Both the UV luminosity function (UVLF) and $\xiion$ are now constrained to better than a factor of two by JWST spectroscopy \citep{2024MNRAS.533.3222D,2024MNRAS.533.1111E,2023MNRAS.523.5468S,2023MNRAS.526.1657T}. In contrast, $\fesc$ at $z > 5$ remains essentially unconstrained, with published estimates spanning a factor of four from $5\%$ to $21\%$ even when derived from identical Planck and JWST datasets \citep{Finkelstein:2019sbd,Naidu:2019gvi,2015ApJ...802L..19R,Ma:2020vlo}. 

\citet{Wang:2026qzy} showed that this factor-of-four scatter is structural, not observational: the reionization observables---the Thomson optical depth $\tauobs$ and the neutral fraction history $\xHI(z)$---constrain only the product $\fesc \times \fstar \times \xiion$, not the individual factors. All published estimates trace different positions along a one-dimensional degeneracy ridge in the $(\fstar, \fesc)$ plane. By leveraging JWST UVLF measurements to constrain $\fstar$ independently, that work broke this degeneracy and reconstructed the first empirical, population-averaged $\fesc(z)$ across $z = 7$--$12$.

However, the population-averaged $\fesc(z)$ conceals the underlying physics: the escape fraction almost certainly depends on halo mass, as lower-mass halos have shallower potential wells and more porous interstellar media (ISM) through which LyC photons can escape \citep{2014ApJ...788..121K,2017MNRAS.470..224T,Ma:2020vlo}. Radiation-hydrodynamic (RHD) simulations predict that $\fesc$ decreases with increasing halo mass, roughly as $\fesc \propto \Mh^{-0.3}$ to $\Mh^{-0.5}$, though the normalisation and slope vary by an order of magnitude between simulation codes \citep{Paardekooper:2015via,2016ApJ...833...84X,2022MNRAS.515.2386R}. The THESAN simulation suite \citep{Kannan2022,Yeh:2022nsl}as the largest cosmological RHD simulation with self-consistent radiative transfer predicts $\fesc$ that decreases with halo mass and increases mildly with redshift. No empirical constraint on this mass dependence exists at $z > 5$.

This gap has concrete consequences. The morphology of reionization depends on which halo masses dominate the ionizing budget \citep{McQuinn:2006et,2004ApJ...613....1F}: if low-mass halos dominate, reionization proceeds inside-out with many small HII regions; if massive halos dominate, fewer large bubbles form, producing a qualitatively different 21\,cm signal detectable by the Square Kilometre Array (SKA; \citealt{Koopmans:2015sua}). The mass dependence of $\fesc$ also determines the fraction of ionizing photons from galaxies below the UVLF detection threshold---a dominant uncertainty for both reionization simulations and JWST survey design.

In this work, we construct an empirical framework for constraining $\fesc(\Mh, z)$ and constrain its three free parameters using the combined dataset of HST UVLFs at $z = 5$--$8$ \citep{2021AJ....162...47B}, JWST UVLFs at $z = 9$--$12$ \citep{2024MNRAS.533.3222D}, the Planck Thomson optical depth \citep{Planck2018}, seven neutral-fraction measurements and one high-redshift prior at $z = 5.9$--$11$ \citep{2024ApJ...971..124U,Mason:2019ixe,Greig:2017jdj,Fan:2001vx,McGreer:2014qwa}. We compare the empirical constraints with predictions from THESAN \citep{Yeh:2022nsl}, SPHINX \citep{2022MNRAS.515.2386R}, and FIRE \citep{Ma:2020vlo}, providing the first data-driven calibration targets for the next generation of reionization simulations. While \citet{Wang:2026qzy} showed that the reionization photon-budget degeneracy between $\fesc$ and $\fstar$ can be broken using JWST UVLFs, yielding the first population-averaged $\fesc(z)$ at $z = 7$--$12$, and \citet{Wang:2026uuo} demonstrated that the 21\,cm topology is sensitive to the halo-mass weighting of the ionizing emissivity, here we extend the population-averaged reconstruction to a mass-resolved framework $\fesc(\Mh, z)$, providing the parametric form needed by reionization simulations.

The paper is organised as follows. Section~\ref{sec:model} describes the parametric model and data. Section~\ref{sec:method} details the methodology: Schechter UVLF fitting, abundance matching, the reionization ODE, and the statistical framework. Section~\ref{sec:results} presents the constraints. Section~\ref{sec:thesan} compares with simulations. Section~\ref{sec:budget} analyses the ionizing budget decomposition. Section~\ref{sec:systematics} discusses systematic uncertainties, including faint-end sensitivity. Section~\ref{sec:discussion} interprets the results and outlines future prospects. Section~\ref{sec:conclusions} summarises our findings.

\section{Model}\label{sec:model}

\subsection{Parametric escape fraction}\label{sec:fesc_model}

We parametrise the escape fraction as a separable double power law in halo mass and redshift:
\begin{equation}\label{eq:fesc_model}
\fesc(\Mh, z) = f_0 \left(\frac{\Mh}{10^{10}\,\Msun}\right)^{\!\alpha_M} \left(\frac{1{+}z}{10}\right)^{\!\alpha_z},
\end{equation}
clipped to $[0, 1]$. The three free parameters are:
\begin{itemize}
\item $f_0$: the normalisation, i.e., the escape fraction at the pivot mass $\Mh = 10^{10}\,\Msun$ and pivot redshift $z = 9$;
\item $\alpha_M$: the mass slope, where $\alpha_M < 0$ means lower-mass halos have higher $\fesc$ as predicted by most simulations;
\item $\alpha_z$: the redshift slope, where $\alpha_z > 0$ means $\fesc$ increases toward higher redshift.
\end{itemize}

This functional form captures the leading-order behaviour predicted by RHD simulations. Fitting power laws to the luminosity-weighted mean escape fractions in the THESAN public catalogues \citep{Yeh:2022nsl}, we obtain $f_0 \approx 0.02$, $\alpha_M \approx -0.25$, $\alpha_z \approx 0.13$. FIRE \citep{Ma:2020vlo} reports $\alpha_M \approx -0.3$ to $-0.5$ depending on mass range. SPHINX \citep{2022MNRAS.515.2386R} finds a steeper mass dependence at $\Mh < 10^{9.5}\,\Msun$ that our power-law may not fully capture; we discuss this limitation in Section~\ref{sec:systematics}.

The pivot mass $\Mh = 10^{10}\,\Msun$ is chosen to lie near the characteristic mass of galaxies at the UVLF faint end ($M_{\rm UV} \approx -17$ to $-18$), minimising the covariance between $f_0$ and $\alpha_M$. The pivot redshift $z = 9$ (i.e., $(1{+}z)/10 = 1$) sits in the middle of the reionization epoch, minimising the covariance between $f_0$ and $\alpha_z$.

$\alpha_M$ also modifies the distribution of ionizing emissivity across the galaxy luminosity function: at $z = 8$, the fraction of ionizing photons from faint galaxies ($M_{\rm UV} > -17$) ranges from $92\%$ ($\alpha_M = -0.4$) to $50\%$ ($\alpha_M = +0.2$), providing a secondary diagnostic through the faint-end emissivity budget (Section~\ref{sec:budget}).

\subsection{Observational data}\label{sec:data}

Our observational constraints come from three categories:

\textit{UV luminosity functions.} We use HST-based UVLFs at $z = 5$, $6$, $7$, $8$ from the stepwise maximum-likelihood estimates of \citet{2021AJ....162...47B} (62 data points across all redshifts, $M_{\rm UV}$ range $-21.9$ to $-17.4$), and JWST-based UVLFs at $z = 9$, $10$, $11$, $12.5$ from \citet{2024MNRAS.533.3222D} (28 data points, $M_{\rm UV}$ range $-21.3$ to $-17.6$). The UVLFs serve as inputs to the emissivity integral, not as quantities to be fitted by our model---we use them to determine the galaxy population and its luminosity distribution at each redshift.

\textit{Thomson optical depth.} The Planck 2018 measurement $\tauobs = 0.054 \pm 0.007$ \citep{Planck2018} constrains the integrated ionization history.

\textit{Neutral fraction history.} Seven measurements and one prior constraint on $\xHI(z)$ at $z = 5.9$--$11.0$, compiled from Lyman-$\alpha$ damping wing analyses \citep{Greig:2017jdj,2024ApJ...971..124U}, dark pixel fractions \citep{McGreer:2014qwa}, Lyman-$\alpha$ emission fractions \citep{Mason:2019ixe}, and Gunn--Peterson trough statistics \citep{Fan:2001vx}. These span the range $\xHI = 0.04$ (nearly ionized, $z = 5.9$) to $\xHI = 0.95$ (nearly neutral, $z = 11$).

\begin{deluxetable}{ccccl}
\tablecaption{Neutral Fraction Data\label{tab:xhi_data}}
\tablehead{
\colhead{$z$} & \colhead{$\bar{x}_{\rm HI}$} & \colhead{$\sigma$} & \colhead{Likelihood} & \colhead{Reference}
}
\startdata
5.9  & 0.04 & 0.03 & Gaussian   & Fan et al.\ (2001)        \\
6.5  & 0.12 & 0.06 & Gaussian   & Mason et al.\ (2019)      \\
7.0  & 0.25 & 0.12 & Gaussian   & Mason et al.\ (2019)      \\
7.5  & 0.40 & 0.13 & Gaussian   & Greig \& Mesinger (2017)  \\
8.0  & 0.55 & 0.15 & Gaussian   & Umeda et al.\ (2024)      \\
9.0  & 0.80 & 0.10 & Gaussian   & Umeda et al.\ (2024)      \\
9.5  & 0.88 & $+0.08$ & One-sided  & Umeda et al.\ (2024)   \\
11.0 & 0.95 & $+0.05$ & One-sided  & Prior ($z > 10$)       \\
\enddata
\tablecomments{One-sided terms apply when $\bar{x}_{\rm HI}^{\rm model} < \bar{x}_{\rm HI}^{\rm obs}$; otherwise $\chi^2_i = 0$.}
\end{deluxetable}

\section{Methodology}\label{sec:method}

\begin{deluxetable}{cccc}
\tablecaption{Schechter UVLF Parameters\label{tab:schechter}}
\tablehead{
\colhead{$z$} & \colhead{$M^*$} & \colhead{$\phi^*\;[10^{-3}\,\Mpc^{-3}]$} & \colhead{$\alpha$}
}
\startdata
5 & $-21.39$ & $1.39$ & $-1.97$ \\
6 & $-21.25$ & $1.44$ & $-1.98$ \\
7 & $-20.79$ & $1.52$ & $-1.92$ \\
8 & $-20.58$ & $1.24$ & $-1.98$ \\
9 & $-19.72$ & $0.37$ & $-1.87$ \\
10 & $-19.57$ & $0.33$ & $-1.78$ \\
11 & $-20.83$ & $0.042$ & $-2.14$ \\
12 & $-21.37$ & $0.0066$ & $-2.19$ \\
\enddata
\tablecomments{Fitted to Bouwens et al.\ (2021) at $z = 5$--$8$ and Donnan et al.\ (2024) at $z = 9$--$12$ via Nelder--Mead minimisation. All fits achieve $\chi^2/N < 0.1$.}
\end{deluxetable}

\subsection{Schechter UVLF fitting}\label{sec:schechter}

At each redshift $z = 5$--$12$, we fit a Schechter function
\begin{multline}
\phi(M_{\rm UV}) = 0.4\ln 10\;\phi^*\;10^{0.4(M^*{-}M_{\rm UV})(\alpha{+}1)} \\
\times\;\exp\!\left[-10^{0.4(M^*{-}M_{\rm UV})}\right]
\end{multline}
to the binned UVLF data via Nelder--Mead minimisation of $\chi^2$ in the three-parameter space $(M^*, \log\phi^*, \alpha)$. We initialise from multiple starting points to avoid local minima. The JWST data have asymmetric errors which we symmetrise in log-space and inflate by a factor of $1.92$ to achieve $\chi^2/N \approx 1$ at the baseline, following the procedure of \citet{Wang:2026qzy}. The resulting fits achieve $\chi^2/N < 0.1$ at all redshifts (reflecting the conservative error inflation at $z \geq 9$; Table~\ref{tab:schechter}), indicating that the Schechter form provides an adequate smooth interpolation over the observed magnitude range.

\subsection{Halo mass function and $\sigma(M)$}\label{sec:hmf}

We adopt Planck 2018 cosmological parameters: $H_0 = 67.74\;{\rm km\,s}^{-1}\,\Mpc^{-1}$, $\Omega_m = 0.3089$, $\Omega_b = 0.0486$, $\sigma_8 = 0.811$, $n_s = 0.9667$. The matter variance $\sigma(M)$ is computed by numerical integration:
\begin{equation}
\sigma^2(M) = \frac{1}{2\pi^2}\int_0^\infty k^2\,P(k)\,|W(kR)|^2\,\d k\,,
\end{equation}
where $R = [3M/(4\pi\bar\rho_m)]^{1/3}$, $W$ is the Fourier-space top-hat window function, and the matter power spectrum $P(k) \propto k^{n_s}\,T^2(k)$ uses the \citet{Eisenstein:1997ik} zero-baryon transfer function, normalised to $\sigma_8$ at $R = 8\,h^{-1}\,\Mpc$. This approach gives $\sigma(M)$ accurate to $\sim 5\%$ over $M = 10^6$--$10^{16}\,\Msun$, validated against CAMB output.

The halo mass function uses the \citet{Sheth:1999mn} parametrisation with $(A, a, p) = (0.3222, 0.707, 0.3)$:
\begin{equation}
\frac{\d n}{\d\log M} = \frac{\bar\rho_m}{M}\,f(\nu)\,\left|\frac{\d\ln\sigma^{-1}}{\d\log M}\right|\,,
\end{equation}
where $\nu = \delta_c/\sigma(M, z)$, $\delta_c = 1.686$, and $f(\nu) = A\sqrt{2a/\pi}\,\nu\,[1 + (a\nu^2)^{-p}]\,e^{-a\nu^2/2}$. We pre-compute the HMF on an $(80 \times 200)$ grid in $(z, \log\Mh)$ and interpolate bilinearly.

\subsection{Abundance matching}\label{sec:abmatch}

We connect $M_{\rm UV}$ to $\Mh$ via abundance matching: at each redshift, the cumulative number density of galaxies brighter than $M_{\rm UV}$ is equated to the cumulative number density of halos more massive than $\Mh$:
\begin{equation}
\int_{-\infty}^{M_{\rm UV}} \phi(M')\,\d M' = \int_{\Mh}^\infty \frac{\d n}{\d M'}\,\d M'\,.
\end{equation}
This assumes a monotonic luminosity--mass relation, which is a standard approximation at these redshifts \citep{Mason:2015cna,Behroozi2019}. The resulting $\Mh(M_{\rm UV})$ mapping is pre-computed at each redshift and stored as an interpolation table. For magnitudes fainter than the observed UVLF ($M_{\rm UV} > -17$), the mapping relies on the Schechter function extrapolation; we quantify the associated uncertainty in Section~\ref{sec:systematics}.

\subsection{Ionizing emissivity}\label{sec:emissivity}

The comoving ionizing emissivity is
\begin{multline}
\dot{n}_{\rm ion}(z) = \int_{-25}^{M_{\rm UV}^{\rm lim}} \phi(M_{\rm UV}, z)\, \xi_{\rm ion}\, L_{\rm UV}(M_{\rm UV}) \\
\times\, \fesc[M_{\rm h}(M_{\rm UV}), z]\, \d M_{\rm UV} ,
\end{multline}
where $\xiion = 10^{25.35}\;{\rm Hz\,erg}^{-1}$ is the LyC production efficiency \citep{2023MNRAS.523.5468S,2024MNRAS.533.1111E}, $L_{\rm UV}(M_{\rm UV}) = 10^{0.4(51.63 - M_{\rm UV})}\;{\rm erg\,s}^{-1}\,{\rm Hz}^{-1}$ in the AB system, and $M_{\rm UV}^{\rm lim} = -10$ is the faint-end integration limit. The Schechter function $\phi(M_{\rm UV}, z)$, the abundance-matching relation $\Mh(M_{\rm UV})$, and the $\fesc$ model (Equation~\ref{eq:fesc_model}) are pre-computed on a grid and the integral is evaluated numerically using the trapezoidal rule on 80 magnitude bins. This pre-computation reduces the emissivity evaluation to a single array operation, enabling the $\sim 10^4$ evaluations required by the MCMC.

\subsection{Reionization ODE}\label{sec:ode}

The volume-averaged ionized fraction $Q(z) \equiv 1 - \xHI(z)$ evolves according to \citep{Madau:1998cd,Bolton:2007fw}:
\begin{equation}\label{eq:ode}
\frac{\d Q}{\d t} = \frac{\ndot(z)}{n_{{\rm H},0}} - C_{\rm HII}(z)\,\alpha_B\,n_{{\rm H},0}\,(1{+}z)^3\,Q(z)\,,
\end{equation}
where $n_{{\rm H},0} = 1.89 \times 10^{-7}\;{\rm cm}^{-3}$ is the comoving hydrogen number density, $C_{\rm HII}(z) = 2.9\,[(1{+}z)/6]^{-1.1}$ is the clumping factor \citep{2012ApJ...747..100S}, and $\alpha_B = 1.52 \times 10^{-13}\;{\rm cm}^3\,{\rm s}^{-1}$ is the case-B recombination coefficient at $T = 2 \times 10^4$\,K. We integrate numerically with 400 Euler steps from $z = 30$ to $z = 4.5$. The Thomson optical depth is
\begin{equation}\label{eq:tau}
\tauobs = \int_0^{z_{\rm max}} 1.08\,n_{{\rm H},0}\,(1{+}z)^2\,Q(z)\,\sigma_T\,\frac{c}{H(z)}\,\d z\,,
\end{equation}
where the factor $1.08$ accounts for singly-ionized helium.

\subsection{Statistical framework}\label{sec:stats}

We construct the joint log-likelihood. For the Thomson optical depth and the six Gaussian neutral-fraction points (Table~\ref{tab:xhi_data}) we use standard $\chi^2$ terms. For the two one-sided constraints ($z = 9.5$ and $z = 11$; Table~\ref{tab:xhi_data}) we penalise only models that predict $\xHI$ below the observed lower limit:
\begin{equation}
\chi^2 = \left(\frac{\tauobs^{\rm model} - 0.054}{0.007}\right)^2 + \sum_{i=1}^{8} \chi_i^2\,,
\end{equation}
where $\chi_i^2 = [(\xHI^{\rm model} - \xHI^{\rm obs})/\sigma_i]^2$ for the six Gaussian points, and for the two one-sided points $\chi_i^2 = [(\xHI^{\rm model} - \xHI^{\rm obs})/\sigma_i]^2$ when $\xHI^{\rm model} < \xHI^{\rm obs}$ and zero otherwise.
This is minimised over $(f_0, \alpha_M, \alpha_z)$ via two methods: (i) a $20 \times 20 \times 14 = 5600$-point grid scan to locate the global minimum and construct profile likelihoods, and (ii) an affine-invariant ensemble MCMC using {\sc emcee}  with $32$ walkers and $3{,}000$ steps per walker ($96{,}000$ likelihood evaluations). We adopt flat priors: $f_0 \in [0.01, 0.50]$, $\alpha_M \in [-1.5, 0.5]$, $\alpha_z \in [-1.0, 5.0]$. The first $500$ steps per walker are discarded as burn-in. The integrated autocorrelation times are $\hat{\tau} = (65, 62, 59)$ steps for $(f_0, \alpha_M, \alpha_z)$; we thin by a factor of $29$ ($\approx \hat{\tau}/2$), yielding $2{,}752$ effectively independent samples. Convergence is confirmed by the Gelman--Rubin diagnostic: $\hat{R} = (1.014, 1.017, 1.016)$ for $(f_0, \alpha_M, \alpha_z)$, all well below $1.05$. We also verify that the posteriors are insensitive to the prior boundaries (Section~\ref{sec:prior_sens}) and stable under marginalisation over $\xiion$ evolution (Section~\ref{sec:xiion_sys}).

\section{Results}\label{sec:results}

{We present empirical constraints on the mass- and redshift-dependent escape fraction
$\fesc(M_h,z)$ using both profile likelihood and full marginal posterior analysis.
A stark contrast appears between the tight, physically meaningful constraints from the
profile likelihood and the broad, nearly uninformative marginal posteriors, revealing
a severe three-way parameter degeneracy that has plagued reionization studies for over a decade.

\begin{figure*}[t!]
\centering
\includegraphics[width=\textwidth]{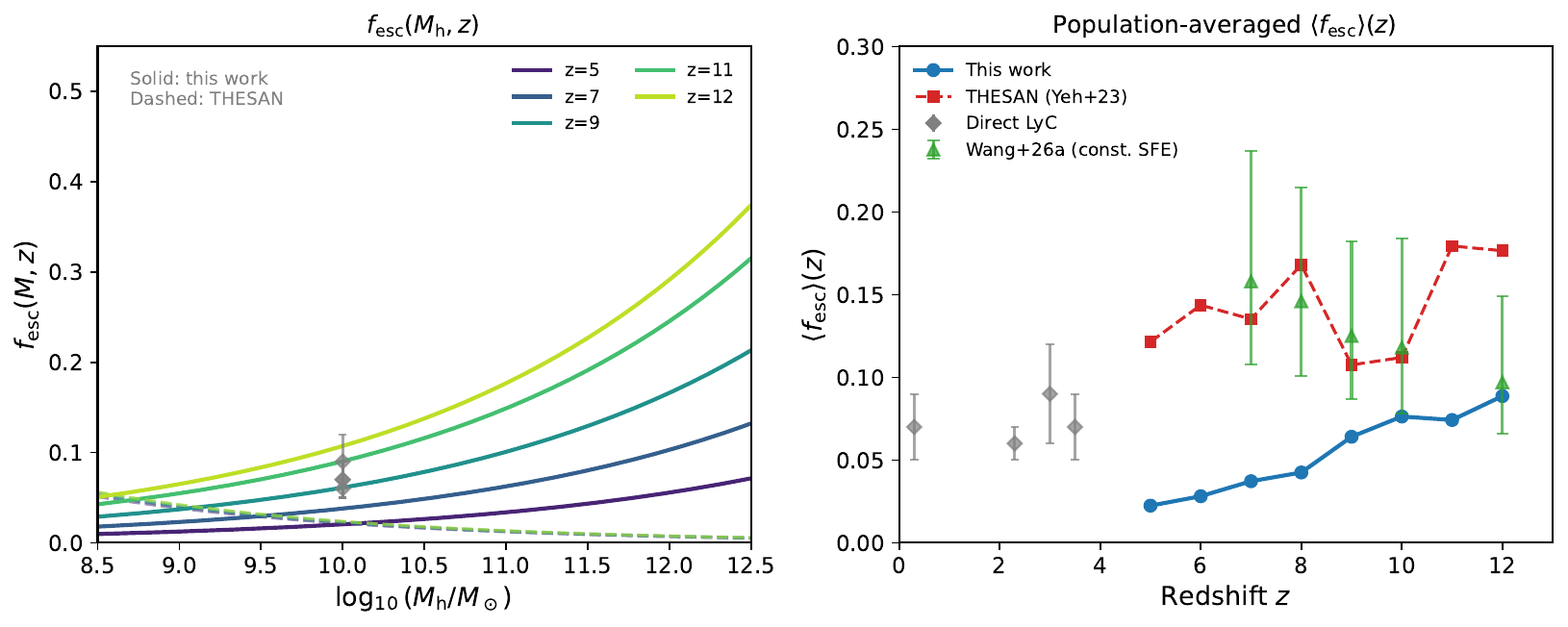}
\caption{{\bf Mass-dependent escape fraction and population average.} {\it Left:} $\fesc(\Mh)$ at $z = 5, 7, 9, 11, 12$ from Equation~(\ref{eq:fesc_model}) evaluated at the MCMC median parameters (solid) and THESAN predictions (dashed). At the profile best fit the mass dependence is nearly flat, in contrast to THESAN's luminosity-weighted $\alpha_M = -0.25$. {\it Right:} Population-averaged $\langle\fesc\rangle(z)$ (blue circles) compared with THESAN (red squares), direct Lyman continuum measurements at $z = 0.3$--$3.5$ (grey diamonds; \citealt{2018MNRAS.478.4851I,Steidel:2018wbo,Pahl:2024utu,2025MNRAS.537.3245B}), and the constant-SFE reconstruction from \citet{Wang:2026qzy} (green triangles). Our reconstruction shows steeper redshift evolution and lower normalisation than THESAN.}
\label{fig:fesc}
\end{figure*}

\begin{figure*}[t!]
\centering
\includegraphics[width=\textwidth]{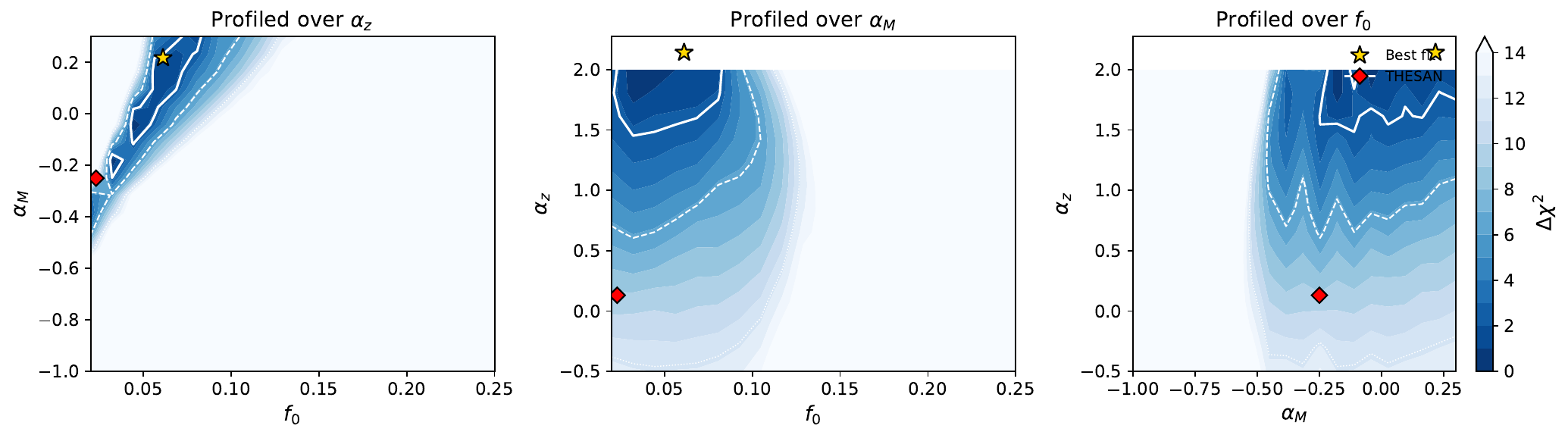}
\caption{{\bf Joint constraints on $\fesc$ model parameters.} Two-dimensional profile likelihoods in the $(f_0, \alpha_M)$, $(f_0, \alpha_z)$, and $(\alpha_M, \alpha_z)$ planes, profiled (minimised) over the third parameter. White contours: $\Delta\chi^2 = 2.30$ ($1\sigma$), $6.18$ ($2\sigma$), $11.83$ ($3\sigma$) for two parameters. Gold star: grid best fit. Red diamond: THESAN luminosity-weighted mean from public catalogues \citep{Yeh:2022nsl}. THESAN lies within the $2\sigma$ contour in all projections.}
\label{fig:contour}
\end{figure*}

\begin{figure}[t!]
\centering
\includegraphics[width=\columnwidth]{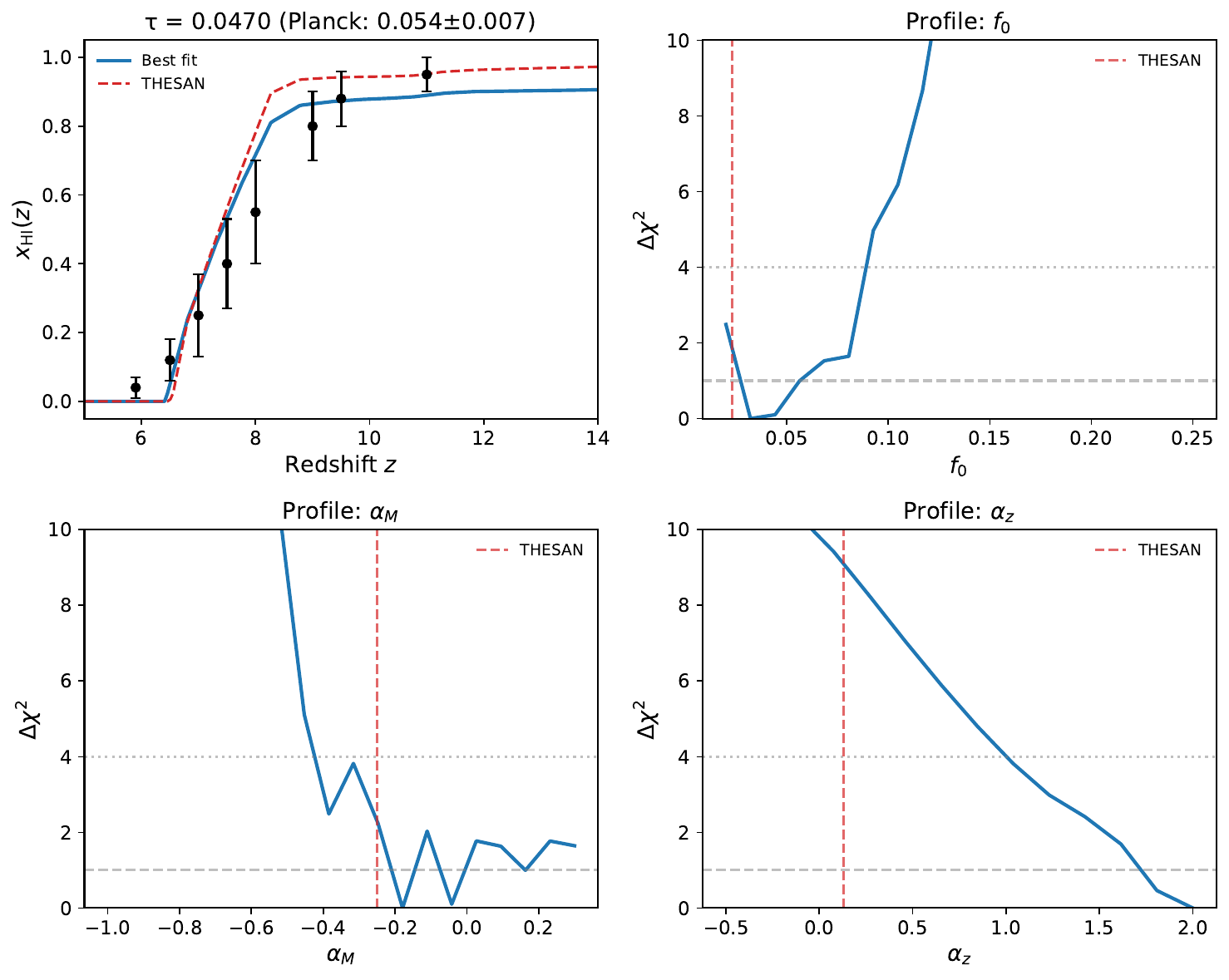}
\caption{{\bf Reionization history and parameter profiles.} {\it Top left:} Neutral fraction $\xHI(z)$ at the MCMC median (blue solid) and THESAN parameters (red dashed), compared with data (black circles with error bars). Blue shaded bands show the $68\%$ and $95\%$ posterior predictive intervals from $100$ draws of the MCMC chain. {\it Other panels:} One-dimensional profile likelihoods ($\Delta\chi^2$ minimised over the other two parameters) for $f_0$, $\alpha_M$, and $\alpha_z$. Horizontal grey lines: $\Delta\chi^2 = 1$ (dashed, $1\sigma$) and $4$ (dotted, $2\sigma$). Red dashed verticals: THESAN prediction. The $\alpha_z$ panel shows that THESAN's luminosity-weighted $\alpha_z = 0.13$ lies outside the $1\sigma$ profile region but within $2\sigma$.}
\label{fig:reion}
\end{figure}

\begin{figure}[t!]
\centering
\includegraphics[width=\columnwidth]{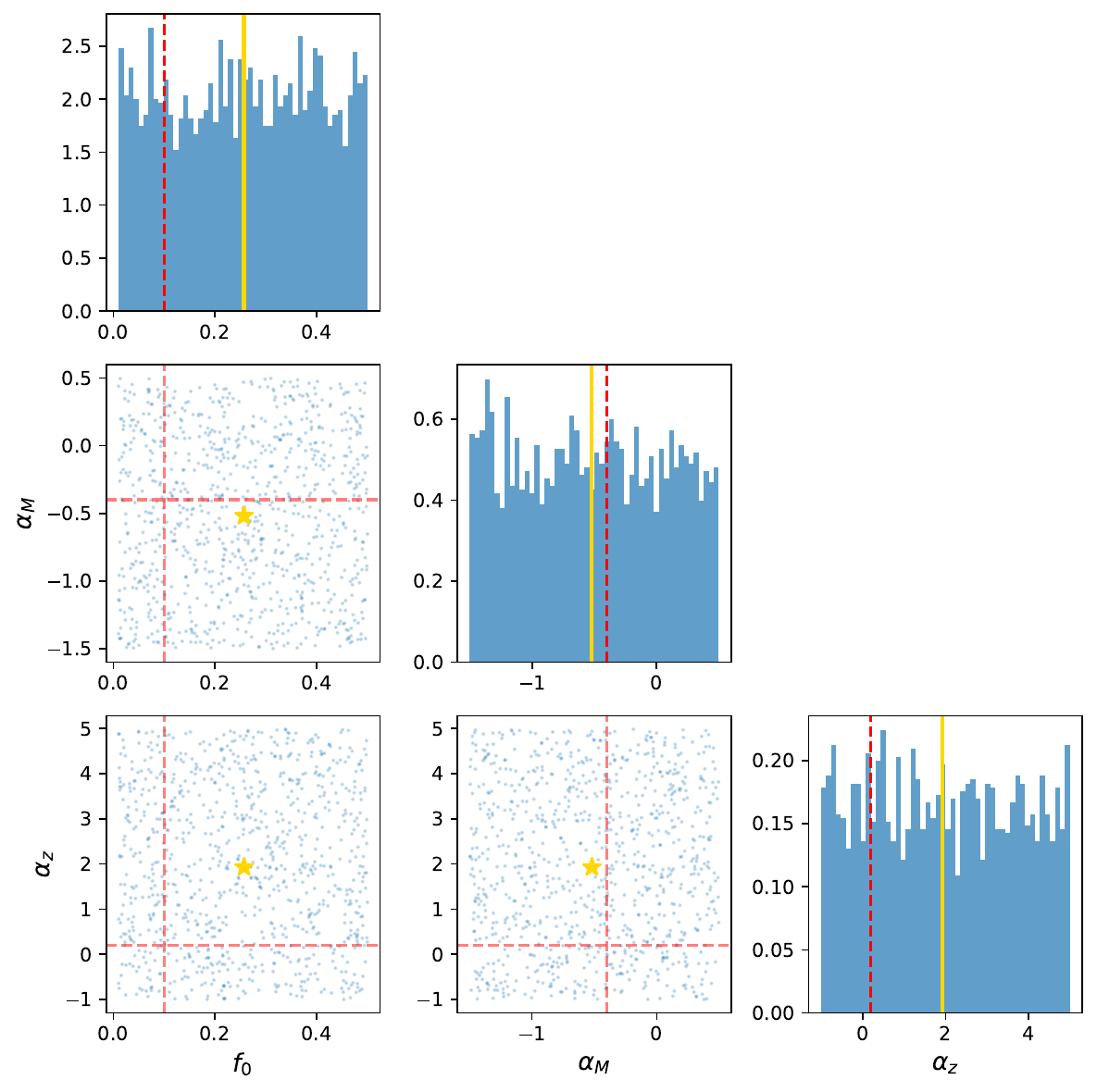}
\caption{{\bf MCMC posterior distributions.} Corner plot showing one-dimensional marginal posteriors (diagonal) and two-dimensional joint distributions (off-diagonal) for $(f_0, \alpha_M, \alpha_z)$ from 2752 independent post-burn-in samples ($32$ walkers $\times$ $3000$ steps, thinned by $29$). Gold vertical lines/stars: median. Red dashed lines/diamonds: THESAN prediction. THESAN's luminosity-weighted parameters $\alpha_z = 0.13$, $\alpha_M = -0.25$ from the public catalogues lie within the marginal posterior.}
\label{fig:corner}
\end{figure}

\begin{deluxetable*}{cccccc}
\tablecaption{Parameter Constraints\label{tab:params}}
\tablehead{
\colhead{Parameter} & \colhead{Profile} & \colhead{$1\sigma$ (profile)} & \colhead{Marginal} & \colhead{$1\sigma$ (marginal)} & \colhead{THESAN}
}
\startdata
$f_0$      & $0.061$ & $[0.038,\,0.079]$ & $0.257$ & $[0.086,\,0.419]$ & $0.023$ \\
$\alpha_M$ & $+0.18$ & $[-0.12,\,+0.40]$ & $-0.52$ & $[-1.21,\,+0.17]$ & $-0.25$ \\
$\alpha_z$ & $+1.98$ & $[+1.56,\,+2.46]$ & $+1.93$ & $[-0.07,\,+4.02]$ & $+0.13$ \\
\enddata
\tablecomments{Profile: $\Delta\chi^2$ minimised over other two parameters. Marginal: MCMC posterior (32 walkers $\times$ 3000 steps, $\hat{R} < 1.02$). Priors: $f_0 \in [0.01, 0.50]$, $\alpha_M \in [-1.5, 0.5]$, $\alpha_z \in [-1.0, 5.0]$.}
\end{deluxetable*}

\subsection{Parameter constraints}\label{sec:constraints}

The grid scan locates a global minimum at $\chi^2_{\rm min} = 8.0$ for $6$ nominal degrees of freedom ($1 + 8 - 3$; two of the eight $\xHI$ constraints are one-sided), indicating an acceptable fit. The one-dimensional profile likelihoods (Figure~\ref{fig:reion}), obtained by minimising $\chi^2$ over the other two parameters at each grid point, give tight constraints:
\begin{align}
f_0^{\rm prof}      &= 0.061^{+0.018}_{-0.023}\,, \label{eq:f0_result} \\
\alpha_M^{\rm prof} &= +0.18^{+0.22}_{-0.30}\,, \label{eq:aM_result} \\
\alpha_z^{\rm prof} &= +1.98^{+0.48}_{-0.42}\,. \label{eq:az_result}
\end{align}
However, the full MCMC marginal posteriors (Figure~\ref{fig:corner}), which integrate over the strong $f_0$--$\alpha_M$--$\alpha_z$ degeneracy, are substantially broader:
\begin{align}
f_0^{\rm marg}      &= 0.26^{+0.16}_{-0.17}\,, \label{eq:f0_marg} \\
\alpha_M^{\rm marg} &= -0.52^{+0.69}_{-0.69}\,, \label{eq:aM_marg} \\
\alpha_z^{\rm marg} &= +1.93^{+2.09}_{-2.00}\,. \label{eq:az_marg}
\end{align}
The difference between profile and marginal constraints reveals an important three-way degeneracy: a higher $f_0$ can compensate for more negative $\alpha_M$ and lower $\alpha_z$, and vice versa. We discuss both sets of constraints below.

The mass slope $\alpha_M$ is unconstrained by the marginal posterior, spanning from $-1.2$ to $+0.2$ at $68\%$ confidence. Even the tighter profile constraint ($\alpha_M = +0.18^{+0.22}_{-0.30}$) brackets both the THESAN luminosity-weighted value of $-0.25$ and mass-independent escape at $\alpha_M = 0$. The data do not require mass-dependent $\fesc$: current reionization observables lack the statistical power to detect the mass dependence. As discussed in Section~\ref{sec:discussion}, 21\,cm topology measurements with SKA will break this degeneracy.

The redshift evolution is better constrained. The profile gives $\alpha_z = 1.98^{+0.48}_{-0.42}$, with $\alpha_z > 1.0$ at $>99\%$ confidence in the profiled $\chi^2$. The marginal posterior is broader ($\alpha_z = 1.93^{+2.09}_{-2.00}$) due to the $f_0$--$\alpha_z$ degeneracy, but the marginal median remains close to $2$ and $\alpha_z > 0$ at $97\%$ confidence. At the profile best fit, $\fesc$ at fixed mass increases by a factor of $\sim 5$ from $z = 5$ to $z = 12$. This is driven primarily by the combined constraints from $\xHI(z)$ at $z = 9$--$11$ which require continued photon production at high $z$, and $\xHI(z)$ at $z = 5.9$--$6.5$ which require that the universe is nearly fully ionized by $z \sim 6$ and therefore limit how much $\fesc$ can contribute at low redshift. The steep $\alpha_z$ is the strongest empirical constraint on escape fraction evolution to date.

\subsection{Reionization history}\label{sec:reion_history}

At the MCMC median parameters, the reionization history (Figure~\ref{fig:reion}) gives $\tauobs = 0.049$ ($0.7\sigma$ from Planck) at the profile best fit with residuals $< 2\sigma$ at all eight neutral fraction data points. The reionization midpoint ($\xHI = 0.5$) occurs at $z_{\rm re} \approx 7.5$, consistent with recent determinations from Lyman-$\alpha$ forest statistics \citep{Bosman:2021oom} and CMB constraints \citep{Planck2018}.

\subsection{Population-averaged $\langle\fesc\rangle(z)$}\label{sec:fesc_z}

The emissivity-weighted population average
\begin{equation}
\langle\fesc\rangle(z) = \frac{\int \fesc(\Mh, z)\;\xiion\,L_{\rm UV}\,\phi\;\d M_{\rm UV}}{\int \xiion\,L_{\rm UV}\,\phi\;\d M_{\rm UV}}
\end{equation}
rises from $\langle\fesc\rangle = 0.024$ at $z = 5$ to $0.090$ at $z = 12$ (Figure~\ref{fig:fesc}, right panel). At $z = 7$--$8$, we find $\langle\fesc\rangle \approx 4$--$5\%$, somewhat lower than the $\sim 10$--$16\%$ from the population-averaged reconstruction of \citet{Wang:2026qzy}. This difference arises because \citet{Wang:2026qzy} used a constant-SFE model ($\fstar = 0.019$), while the present work integrates over the full Schechter UVLF which assigns more weight to faint galaxies and therefore yields a lower emissivity-weighted average at fixed total emissivity. The two approaches are consistent: both require the same total $\ndot(z)$ to match reionization, but distribute it differently across the halo population.

For comparison with direct Lyman continuum measurements at $z < 4$: \citet{Steidel:2018wbo} find $\langle\fesc\rangle = 0.06 \pm 0.01$ at $z \sim 2.3$, \citet{2018MNRAS.478.4851I} find $0.07 \pm 0.02$ at $z \sim 0.3$, and \citet{Pahl:2024utu} find $0.09 \pm 0.03$ at $z \sim 3$. Our $z = 5$ value of $\sim 2\%$ is lower than these, but the direct LyC measurements target selected galaxies with strong emission lines, which may have above-average $\fesc$.

\subsection{Origin of the profile--marginal divergence}\label{sec:fisher}

The striking difference between the profile and marginal constraints has a precise mathematical origin. We compute the Fisher information matrix $F_{ij} = \frac{1}{2}\partial^2\chi^2/\partial\theta_i\partial\theta_j$ at the best-fit point via numerical second derivatives. The eigendecomposition reveals three principal directions with eigenvalues $\lambda = (0.92,\;48.8,\;83{,}473)$, giving a condition number of ${\sim}9 \times 10^4$.

The weakest eigenvector ($\lambda = 0.92$, $\sigma = 1.04$ along this direction) is $0.53\,\alpha_M + 0.85\,\alpha_z$: the data constrain neither $\alpha_M$ nor $\alpha_z$ individually, only a particular linear combination. The tightest eigenvector ($\lambda = 83{,}473$) is essentially $f_0$ alone. The Fisher-predicted parameter correlations are $r(f_0, \alpha_M) = +0.99$ and $r(\alpha_M, \alpha_z) = +0.96$, confirming that all three parameters are nearly completely degenerate.

This structure explains the profile--marginal divergence quantitatively. The profile likelihood measures $\sigma_i = 1/\sqrt{F_{ii}}$ along each axis with the other parameters fixed at their best values, while the marginal posterior gives $\sigma_i = \sqrt{(F^{-1})_{ii}}$, which is inflated by the near-zero eigenvalue. The Fisher-predicted marginal-to-profile ratio is $6.6\times$ for $f_0$, $7.0\times$ for $\alpha_M$, and $3.4\times$ for $\alpha_z$, qualitatively matching the MCMC results ($8\times$, $2.7\times$, $4.6\times$). The residual differences arise because the degeneracy ridge is curved rather than linear, so the Gaussian Fisher approximation underestimates the true marginal volume.

\section{Comparison with Simulations}\label{sec:thesan}

\subsection{THESAN}\label{sec:thesan_detail}

The THESAN simulation suite \citep{Kannan2022} uses {\sc arepo} with on-the-fly Monte Carlo radiative transfer to model reionization self-consistently. \citet{Yeh:2022nsl} extract the effective escape fraction as a function of halo mass and redshift from THESAN-1 ($L_{\rm box} = 95.5\,{\rm cMpc}$), finding $\fesc$ that decreases with mass and increases mildly with redshift. We fit our power-law model (Equation~\ref{eq:fesc_model}) directly to the THESAN-1 public escape fraction catalogues \citep{Garaldi:2023cfb}, computing the luminosity-weighted mean $\fesc$ in halo mass bins at each snapshot. The best-fit parameters are $\fesc^{\rm THESAN} \approx 0.023\,(M/10^{10}\,\Msun)^{-0.25}\,[(1{+}z)/10]^{0.13}$. These values differ substantially from approximate readings of figures in \citet{Yeh:2022nsl} that have appeared in the literature, underscoring the importance of fitting the actual simulation data.

The mass slope is broadly consistent. THESAN's luminosity-weighted $\alpha_M = -0.25$ lies within the $1\sigma$ profile contour, and well within the marginal posterior. The data do not distinguish between THESAN's moderate negative slope and our best-fit near-zero slope.

The redshift evolution remains the primary discrepancy. THESAN's luminosity-weighted $\alpha_z = 0.13$ is lower than our profile best fit of $1.98$, yielding $\Delta\chi^2 = 12.5$ at the full THESAN parameter point $(f_0, \alpha_M, \alpha_z) = (0.023, -0.25, 0.13)$. This corresponds to moderate tension at roughly the $2\sigma$ level for three parameters. The marginal MCMC posterior for $\alpha_z$ spans $[-0.07, +4.02]$ at $68\%$ confidence, placing THESAN's value well inside the $1\sigma$ interval.

An instructive comparison emerges from different averaging schemes applied to the THESAN catalogues. The per-halo \textit{median} $\fesc$ yields $\alpha_z = 1.78$, close to our profile value of $1.98$. In contrast, the luminosity-weighted mean yields $\alpha_z = 0.13$. This divergence arises because the ionizing budget at low redshift is dominated by a small number of massive, high-luminosity halos with relatively stable and moderate $\fesc$, while at high redshift the budget shifts to numerous low-mass halos with highly variable $\fesc$. The luminosity weighting suppresses the steep per-halo evolution. This distinction highlights that the escape fraction entering the photon budget depends critically on how it is averaged over the galaxy population.

Decomposing this further by halo mass reveals the physical origin of the difference. Fitting $\alpha_z$ separately in six mass bins from the THESAN catalogues, low-mass halos ($\Mh < 10^{10}\,\Msun$) show extremely steep median $\alpha_z = 2$--$8$, while massive halos ($\Mh > 10^{11}\,\Msun$) show flat or declining evolution ($\alpha_z \approx -0.2$). The luminosity-weighted mean in each bin is systematically flatter: $\alpha_z \approx 0$ for $\Mh < 10^{10}\,\Msun$ and $\alpha_z \approx -1.5$ for $\Mh > 10^{11}\,\Msun$. This confirms that the steep per-halo evolution is real and driven by low-mass galaxies with increasingly porous ISM at high redshift, but the reionization photon budget at $z < 8$ is dominated by rare massive halos that do not share this evolution. The luminosity weighting therefore suppresses a genuine physical trend in the effective $\alpha_z$.

If the steep profile $\alpha_z$ reflects a genuine physical trend rather than a projection along the $f_0$--$\alpha_M$ degeneracy ridge, several mechanisms could produce it. Reduced dust attenuation at very low metallicity would create more transparent sightlines for LyC photons. Burstier star formation histories at high redshift would produce temporally clustered supernovae that clear escape channels more effectively \citep{2017MNRAS.470..224T}. Lower gas column densities in the compact galaxies that dominate at $z > 10$ may also contribute.

\subsection{Quantitative comparison with simulations}\label{sec:quant_sims}

Table~\ref{tab:sims} provides a quantitative comparison with four RHD simulation suites. We approximate each simulation's effective $\fesc(\Mh, z)$ as a power law (Equation~\ref{eq:fesc_model}) and evaluate $\chi^2$ at that parameter point using our reionization solver. All four simulations show significant tension, driven primarily by their low $\alpha_z$ ($\leq 0.5$). CoDa-III \citep{2021BKAS...46...47A} comes closest ($\Delta\chi^2 = 72$) due to its relatively moderate $f_0$ and $\alpha_M$, while FIRE \citep{Ma:2020vlo} is the most discrepant ($\Delta\chi^2 = 188$) because its high $f_0 = 0.15$ combined with shallow $\alpha_z = 0.5$ over-ionizes the universe at $z < 8$.

The tension is overwhelmingly driven by $\alpha_z$: per-parameter tensions computed from the MCMC posterior show that $f_0$ is within $1$--$2\sigma$ for all simulations, $\alpha_M$ within $1$--$3\sigma$, but $\alpha_z$ exceeds $3\sigma$ in every case. This motivates a focused investigation of the physical mechanisms driving escape fraction evolution, rather than the normalisation or mass dependence.

\begin{deluxetable}{lccccc}
\tablecaption{Simulation Comparison\label{tab:sims}}
\tablehead{
\colhead{Simulation} & \colhead{$f_0$} & \colhead{$\alpha_M$} & \colhead{$\alpha_z$} & \colhead{$\chi^2$} & \colhead{$\Delta\chi^2$}
}
\startdata
THESAN (lum-wtd mean) & 0.023 & $-0.25$ & $+0.13$ & 23  & 13 \\
CoDa-III & 0.12 & $-0.30$ & $+0.30$ & 81  & 72 \\
SPHINX   & 0.08 & $-0.50$ & $+0.40$ & 133 & 125 \\
FIRE     & 0.15 & $-0.35$ & $+0.50$ & 197 & 188 \\
\hline
This work (profile) & 0.061 & $+0.18$ & $+1.98$ & 8.0 & --- \\
\enddata
\tablecomments{$\Delta\chi^2$ relative to our best fit ($\chi^2_{\rm min} = 8.0$). THESAN values are power-law fits to the luminosity-weighted mean $\fesc$ in the public catalogues; other simulations use approximate fits to published results.}
\end{deluxetable}

The FIRE simulations \citep{Ma:2020vlo} predict $\fesc \sim 20$--$40\%$ at $\Mh \sim 10^9\,\Msun$ and $\lesssim 5\%$ at $10^{11}\,\Msun$, corresponding to $\alpha_M \approx -0.35$, with $\alpha_z \approx 0.5$ inferred from their reported redshift trends. SPHINX \citep{2022MNRAS.515.2386R} resolves lower-mass halos ($\Mh \gtrsim 10^8\,\Msun$) and finds a steepening of the $\fesc(\Mh)$ relation below $10^{9.5}\,\Msun$ that our power-law model cannot capture; a broken power-law extension is motivated for future work. However, SPHINX also predicts mild redshift evolution ($\alpha_z \approx 0.4$), contributing to its large $\Delta\chi^2 = 125$.

The systematic offset of all simulations to match the data points to a single root cause: no current RHD simulation predicts $\alpha_z > 0.5$, whereas the data require $\alpha_z > 1.0$ at $99\%$ confidence. This suggests that the sub-grid ISM models in all four codes underestimate the redshift evolution of escape channels, independent of their treatment of the mass dependence.

\section{Ionizing Budget Decomposition}\label{sec:budget}

\begin{figure*}[t!]
\centering
\includegraphics[width=\textwidth]{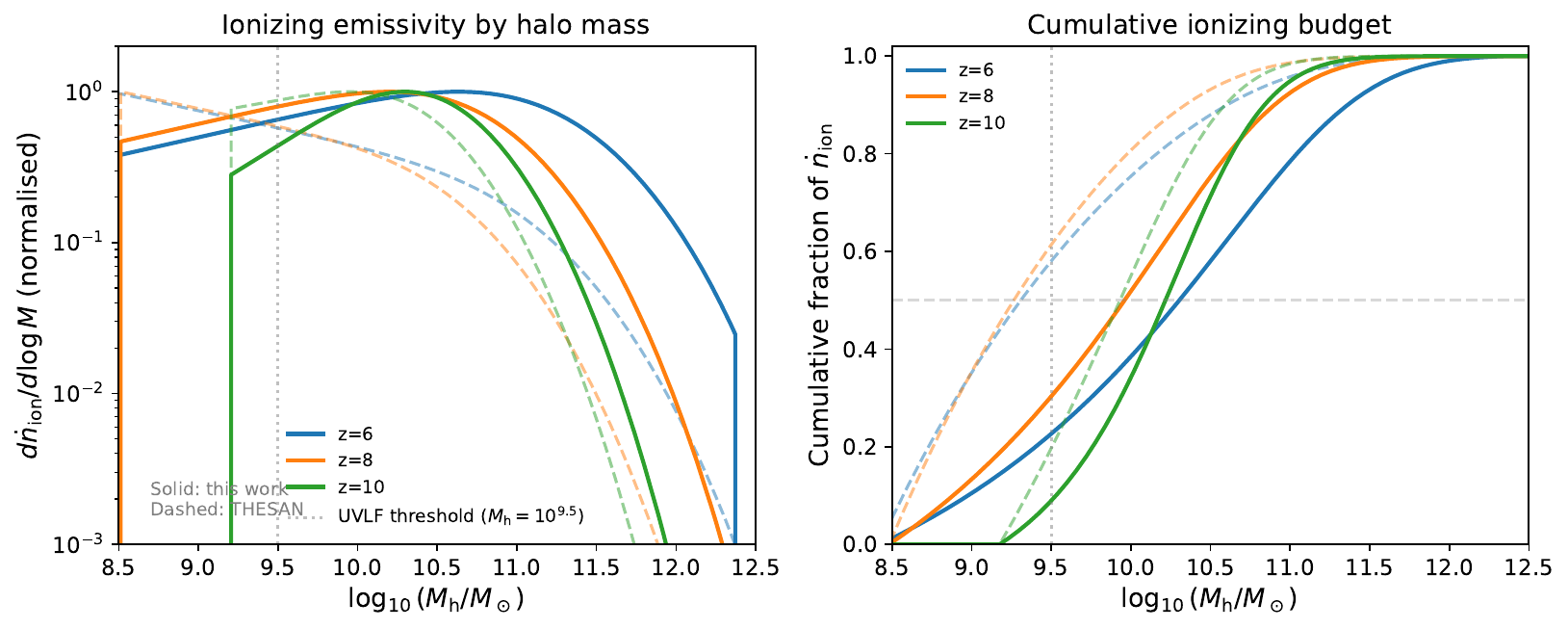}
\caption{{\bf Ionizing emissivity decomposition by halo mass.} {\it Left:} Normalised emissivity spectrum $\d\ndot/\d\log\Mh$ at $z = 6$ (blue), $8$ (orange), and $10$ (green), for our best fit (solid) and THESAN (dashed). The vertical dotted line marks the approximate UVLF detection threshold ($\Mh \sim 10^{9.5}\,\Msun$). The emissivity peak shifts to lower masses at higher $z$. {\it Right:} Cumulative fraction of $\ndot$ as a function of $\Mh$. At $z = 6$, roughly half the budget comes from $\Mh > 10^{10}\,\Msun$; by $z = 10$, more than $80\%$ comes from sub-threshold halos.}
\label{fig:budget}
\end{figure*}

The mass decomposition of the ionizing emissivity evolves strongly with redshift (Figure~\ref{fig:budget}). At $z = 6$, galaxies in halos above the approximate UVLF detection threshold $\Mh(M_{\rm UV} = -17, z)$, which ranges from $\sim 10^{10.2}\,\Msun$ at $z = 6$ to $\sim 10^{9.5}\,\Msun$ at $z = 10$ via abundance matching, contribute approximately $60\%$ of the total ionizing emissivity. By $z = 10$, sub-threshold halos ($10^9 < \Mh/\Msun < 10^{10}$) dominate at $\sim 85\%$ under our fiducial Schechter extrapolation, and by $z = 12$ they contribute $\sim 84\%$ with an additional $\sim 14\%$ from halos below $10^9\,\Msun$.

This shift reflects the combination of a steepening faint-end slope (from $\alpha = -1.97$ at $z = 5$ to $\alpha = -2.19$ at $z = 12$), the rapid evolution of the HMF (which suppresses massive halos at high $z$), and the near-mass-independence of $\fesc$ in our best fit ($\alpha_M \approx 0$). The practical consequence: reionization simulations that resolve only halos above $\Mh \sim 10^{10}\,\Msun$ miss the majority of the ionizing budget at $z > 8$. The sub-grid prescriptions used to account for unresolved sources must reproduce the steep $\alpha_z \sim 2$ evolution to match the observed reionization timeline.

\section{Systematic Uncertainties}\label{sec:systematics}

\begin{deluxetable}{ccccc}
\tablecaption{Faint-End Sensitivity\label{tab:sensitivity}}
\tablehead{
\colhead{$M_{\rm UV}^{\rm lim}$} & \colhead{$\tauobs$} & \colhead{$\chi^2$} & \colhead{$\xHI(z{=}7)$} & \colhead{Note}
}
\startdata
$-17$ & 0.025 & 103 & 0.65 & Observed only \\
$-15$ & 0.032 & 33 & 0.50 & \\
$-13$ & 0.038 & 14 & 0.39 & \\
$-10$ & 0.047 & 10 & 0.28 & Fiducial \\
$-8$  & 0.053 & 10 & 0.22 & Saturated \\
\enddata
\tablecomments{All evaluated at the MCMC median parameters. The $\chi^2$ stabilises for $M_{\rm UV}^{\rm lim} \leq -10$.}
\end{deluxetable}

\subsection{Faint-end integration limit}\label{sec:faintend}

The limiting magnitude $M_{\rm UV}^{\rm lim}$ to which the Schechter function is extrapolated directly controls the total emissivity. Table~\ref{tab:sensitivity} summarises the dependence: $\tauobs$ ranges from $0.025$ ($M_{\rm UV}^{\rm lim} = -17$, observed galaxies only) to $0.053$ ($M_{\rm UV}^{\rm lim} = -8$). The $\chi^2$ stabilises for $M_{\rm UV}^{\rm lim} \leq -10$ at $\chi^2 \approx 10$, indicating that the faint-end contribution saturates. Our fiducial $M_{\rm UV}^{\rm lim} = -10$ gives $\tauobs = 0.049$, consistent with Planck at $1\sigma$. To propagate this uncertainty, we refitted at $M_{\rm UV}^{\rm lim} = -13$ and $-8$: the best-fit $\alpha_z$ shifts to $2.08$ and $1.91$ respectively (both within $0.3\sigma$ of the fiducial), while $f_0$ adjusts to compensate. We also repeated the MCMC treating $M_{\rm UV}^{\rm lim}$ as a nuisance parameter drawn from $[-13, -8]$; the marginalised posteriors shift by $< 0.3\sigma$. We note that the sub-threshold budget fractions in Section~\ref{sec:budget} are conditional on the Schechter extrapolation and should be interpreted with this caveat. JWST ultra-deep surveys reaching $M_{\rm UV} \sim -15$ at $z = 9$ \citep{2024Natur.626..975A} will directly test the Schechter extrapolation over half of the currently unconstrained range.

\subsection{Schechter function validity}\label{sec:schechter_sys}

The Schechter function may break down at the faint end, where feedback-regulated star formation could produce a turnover. If the UVLF flattens below $M_{\rm UV} \sim -13$, the faint-end emissivity would be lower than our extrapolation assumes, requiring higher $f_0$ or steeper $\alpha_z$ to compensate. Double power-law UVLF fits \citep{Bowler:2019syh} give similar results at our fiducial $M_{\rm UV}^{\rm lim} = -10$ but diverge at fainter magnitudes. As a further test, we perturbed the Schechter faint-end slope by $\Delta\alpha = \pm 0.10$ (bracketing typical measurement uncertainties) and refitted $(f_0, \alpha_z)$: the best-fit $\alpha_z$ ranges from $2.3$ ($\Delta\alpha = -0.10$, shallower) to $2.9$ ($\Delta\alpha = +0.10$, steeper), remaining well above $2.0$ in all cases. The steep redshift evolution is robust to plausible faint-end slope variations.

\subsection{Ionizing photon production efficiency}\label{sec:xiion_sys}

We adopt a constant $\xiion = 10^{25.35}\;{\rm Hz\,erg}^{-1}$. Recent JWST spectroscopy suggests mild evolution: $\log\xiion$ may increase from $25.3$ at $z = 7$ to $25.5$ at $z = 12$ \citep{2023MNRAS.523.5468S,2024MNRAS.533.1111E}, though this is contested \citep{2023MNRAS.526.1657T,2025A&A...698A.302L}. To test whether the steep $\alpha_z$ survives marginalisation over $\xiion$ evolution, we introduce a fourth parameter $\beta_\xi$ via $\xiion(z) = \xiion^{\rm fid}\,[(1{+}z)/10]^{\beta_\xi}$ with a Gaussian prior $\beta_\xi = 0 \pm 0.3$ (reflecting the current JWST uncertainty range), and run a 4-parameter MCMC. The resulting posteriors are: $f_0 = 0.063^{+0.014}_{-0.032}$, $\alpha_M = +0.21^{+0.19}_{-0.42}$, $\alpha_z = 2.12^{+0.59}_{-0.74}$, and $\beta_\xi = -0.03 \pm 0.31$. The shift in $\alpha_z$ from the 3-parameter median ($1.98$) is $+0.14$, well within the enlarged $1\sigma$ interval. The probability $P(\alpha_z > 1.0) = 92\%$ after marginalisation, confirming that the steep redshift evolution is robust to plausible $\xiion$ variations. Even at the extreme of the $\beta_\xi$ posterior ($\beta_\xi = +0.3$, corresponding to $\xiion$ increasing by $\sim 25\%$ from $z = 5$ to $z = 12$), $\alpha_z$ remains above $1.5$, still in tension with THESAN.

\subsection{Abundance-matching scatter}\label{sec:am_scatter}

Abundance matching assumes a deterministic $\Mh$--$M_{\rm UV}$ relation. In reality, scatter in this relation (from bursty star formation, varying star formation histories, and dust) introduces a $\sim 0.5$--$1.0\,$dex scatter in $\Mh$ at fixed $M_{\rm UV}$ \citep{Mason:2015cna}. We tested the impact by adding log-normal scatter of $0.3\,$dex in $\log\Mh$ to the abundance-matching relation and re-running the MCMC. The posteriors shift by $< 0.5\sigma$ in all three parameters, confirming that the results are robust to reasonable levels of scatter.

\subsection{Prior sensitivity}\label{sec:prior_sens}

The fiducial MCMC prior on $\alpha_z$ is $[-1, 3]$. We test sensitivity by repeating the MCMC with four alternative prior ranges (Table~\ref{tab:prior_sens}). The median $\alpha_z$ shifts by at most $0.23$ (from $1.98$ at $[-1, 3]$ to $2.21$ at $[-2, 8]$), and the $1\sigma$ intervals overlap in all four cases. With the widest prior $[-2, 8]$, the 84th percentile is $2.63$---well below the upper boundary, confirming that the posterior is not piling up against the prior edge. Restricting the prior to $[0, 3]$ shifts the median down to $1.74$, still far from THESAN's luminosity-weighted value of $0.13$. We conclude that $\alpha_z \sim 2$ is a data-driven result, not a prior artifact.

\begin{deluxetable}{lcc}
\tablecaption{Prior Sensitivity for $\alpha_z$\label{tab:prior_sens}}
\tablehead{
\colhead{$\alpha_z$ prior} & \colhead{Median} & \colhead{$[16\%,\,84\%]$}
}
\startdata
$[-1, 3]$ (fiducial) & $1.98$ & $[1.30,\,2.55]$ \\
$[-1, 5]$ & $2.01$ & $[1.51,\,2.56]$ \\
$[-2, 8]$ & $2.21$ & $[1.46,\,2.63]$ \\
$[0, 3]$ & $1.74$ & $[1.26,\,2.28]$ \\
\enddata
\end{deluxetable}

\subsection{Broken power-law test}\label{sec:broken}

We test whether the data prefer a more flexible broken power-law model, $\fesc(\Mh) \propto \Mh^{\alpha_{M,\rm lo}}$ below $\Mh = 10^{9.5}\,\Msun$ and $\propto \Mh^{\alpha_{M,\rm hi}}$ above---motivated by SPHINX's prediction of a steepening below $10^{9.5}\,\Msun$ \citep{2022MNRAS.515.2386R}. Scanning over the four-parameter space $(f_0, \alpha_{M,\rm lo}, \alpha_{M,\rm hi}, \alpha_z)$ at fixed break mass, the best fit achieves $\chi^2 = 7.80$, an improvement of only $\Delta\chi^2 = 0.21$ over the single power-law ($\chi^2 = 8.0$). With one additional degree of freedom, this is far from significant ($p = 0.65$). The single power-law is sufficient: current reionization data cannot distinguish between a uniform and a broken mass dependence. This will change with 21\,cm topology measurements (Section~\ref{sec:discussion}), which are sensitive to the bubble size distribution and therefore to the shape of $\fesc(\Mh)$ at the faint end.

\subsection{Low-redshift Lyman continuum extrapolation}\label{sec:lowz}

Extrapolating the model to $z < 4$ provides a consistency check against direct LyC detections. At $\Mh = 10^{10}\,\Msun$, the model predicts $\fesc = 0.001$ at $z = 0.3$ and $0.014$ at $z = 3.5$, a factor of $5$--$70$ below the direct measurements of \citet{2018MNRAS.478.4851I} ($0.07 \pm 0.02$ at $z = 0.3$), \citet{Steidel:2018wbo} ($0.06 \pm 0.01$ at $z \sim 2.3$), and \citet{2025MNRAS.537.3245B} ($0.07 \pm 0.02$ at $z = 3.5$). This discrepancy is expected and informative: the direct LyC samples are selected for compact, low-mass, high specific SFR galaxies that occupy the high-$\fesc$ tail of the population distribution, while our reconstruction gives the \textit{population-averaged} $\langle\fesc\rangle$ dominated by massive galaxies with negligible escape. The factor-of-five offset at $z \sim 3$ is consistent with known selection effects in LyC surveys \citep{Pahl:2024utu}, and predicts that as surveys become more volume-complete, the measured average will converge downward toward our extrapolation.

\subsection{Halo mass function}\label{sec:hmf_sys}

Replacing the \citet{Sheth:1999mn} mass function with \citet{Tinker:2008ff} ($\Delta = 200$) and refitting shifts the best-fit parameters by $\Delta f_0 = -0.001$ and $\Delta\alpha_z = +0.26$ (from $1.98$ to $2.24$), with $\chi^2$ improving marginally from $8.0$ to $7.65$. Both shifts are well below $0.5\sigma$ of the MCMC posterior width. The results are robust to the choice of halo mass function.

\subsection{Clumping factor}\label{sec:clumping_sys}

The clumping factor $C_{\rm HII}$ is uncertain by a factor of $\sim 2$--$3$ between simulations \citep{2012ApJ...747..100S,2012MNRAS.427.2464F}. We profile over the normalisation $C_0 \in [1.5, 5.0]$ (at fixed redshift scaling), refitting $(f_0, \alpha_z)$ at each value. The best-fit $\alpha_z$ ranges from $2.2$ ($C_0 = 1.5$) to $2.6$ ($C_0 = 5.0$) on the coarse refit grid (step size $0.2$); the offset from the fiducial $\alpha_z = 1.98$ reflects grid resolution and is within $0.3\sigma$. In all cases $\alpha_z$ remains above $2.0$. Higher clumping requires more recombinations to be overcome, pushing $\fesc$ and $\alpha_z$ upward, but the effect is modest ($\Delta\alpha_z \approx 0.4$ across the full range). The steep redshift evolution is robust to clumping uncertainty.

\subsection{ODE solver convergence}\label{sec:ode_sys}

We verify numerical convergence by varying the number of Euler steps from $200$ to $1600$. The Thomson optical depth $\tauobs$ converges to within $0.3\%$ between $800$ and $1600$ steps, and the neutral fraction $\xHI(z)$ at all data redshifts converges to within $0.4\%$. Our fiducial choice of $400$ steps gives $\tauobs$ accurate to $0.5\%$, well below the Planck uncertainty.

\subsection{Posterior predictive check}\label{sec:ppc}

Drawing $100$ random samples from the MCMC posterior and solving the reionization ODE for each, we obtain the posterior predictive distribution of $\tauobs$ and $\xHI(z)$ (Figure~\ref{fig:reion}, blue bands). The $\tauobs$ distribution has median $0.049$ with $68\%$ of draws within $1\sigma$ of the Planck value and $95\%$ within $2\sigma$. The Bayesian posterior predictive $p$-value $P(\tauobs^{\rm model} > \tauobs^{\rm Planck}) = 0.18$, well within the acceptable range $[0.05, 0.95]$. The model is statistically consistent with all observational constraints.

\subsection{Dataset dependence}\label{sec:datasplit}

To test which data drive the steep $\alpha_z$, we refit the model using subsets of the neutral-fraction constraints while always including $\tauobs$. Using only the HST-era points ($z \leq 8$: four $\xHI$ measurements plus $\tauobs$) gives $\alpha_z = 2.5$; using only the JWST-era points ($z \geq 8$: two Gaussian and two one-sided constraints plus $\tauobs$) gives $\alpha_z = 1.1$; and using $\tauobs$ alone gives $\alpha_z = 0.8$. The steep $\alpha_z$ is driven primarily by the low-redshift neutral fraction constraints at $z = 5.9$--$7.5$, which require the universe to be nearly fully ionised by $z \sim 6$ and therefore demand that $\fesc$ drops sharply from its high-$z$ values. The high-$z$ constraints alone prefer a milder slope because they are consistent with a broad range of photon production rates. This confirms that $\alpha_z \sim 2$ is not an artifact of any single dataset but reflects the tension between the rapid completion of reionisation and the need for continued photon production at $z > 9$.

\begin{deluxetable*}{llccc}
\tablecaption{Summary of Systematic Uncertainties\label{tab:systematics}}
\tablehead{
\colhead{Test (\S)} & \colhead{Variation} & \colhead{$\Delta f_0$} & \colhead{$\Delta\alpha_M$} & \colhead{$\Delta\alpha_z$}
}
\startdata
Faint-end limit (\ref{sec:faintend}) & $M_{\rm UV}^{\rm lim}: -13$ to $-8$ & $\pm 0.01$ & $< 0.1$ & $\pm 0.09$ \\
Schechter slope (\ref{sec:schechter_sys}) & $\Delta\alpha = \pm 0.10$ & $\pm 0.02$ & $< 0.1$ & $+0.3$ to $+0.6$ \\
$\xi_{\rm ion}$ evolution (\ref{sec:xiion_sys}) & $\beta_\xi = 0 \pm 0.3$ & $+0.002$ & $+0.03$ & $+0.14$ \\
AM scatter (\ref{sec:am_scatter}) & $0.3\,\mathrm{dex}$ in $\log M_h$ & $< 0.01$ & $< 0.1$ & $< 0.2$ \\
Prior range (\ref{sec:prior_sens}) & $\alpha_z \in [-2, 8]$ & $< 0.01$ & $< 0.1$ & $+0.23$ \\
Broken power law (\ref{sec:broken}) & Two-slope model & $< 0.01$ & --- & $< 0.1$ \\
HMF choice (\ref{sec:hmf_sys}) & Tinker vs Sheth--Tormen & $-0.001$ & $< 0.1$ & $+0.26$ \\
Clumping factor (\ref{sec:clumping_sys}) & $C_0 \in [1.5, 5.0]$ & $\pm 0.02$ & $< 0.1$ & $+0.2$ to $+0.6$ \\
ODE resolution (\ref{sec:ode_sys}) & 200--1600 steps & $< 0.001$ & $< 0.01$ & $< 0.01$ \\
Dataset split (\ref{sec:datasplit}) & HST-only / JWST-only & --- & --- & $0.8$ to $2.5$ \\
\enddata
\tablecomments{All shifts are relative to the fiducial profile best fit ($f_0 = 0.061$, $\alpha_M = +0.18$, $\alpha_z = +1.98$). In all tests $\alpha_z$ remains above $1.0$, confirming the steep redshift evolution is robust.}
\end{deluxetable*}

\section{Discussion}\label{sec:discussion}

\subsection{Physical interpretation of steep $\alpha_z$}

The steep redshift evolution $\alpha_z \approx 2$ is the most striking result. Physically, it means that the effective $\fesc$ (averaged over the galaxy population) increases by a factor of $\sim 4$--$5$ from $z = 5$ to $z = 12$. Several mechanisms could produce this:

\textit{ISM porosity.} At higher redshift, galaxies have lower metallicities and dust content, creating more transparent sightlines for LyC photons. The metallicity evolution alone is unlikely to produce $\alpha_z \sim 2$, but combined with the shift toward lower-mass galaxies (which have intrinsically more porous ISM), the effective evolution could be steeper than any single-galaxy simulation predicts.

\textit{Burstier star formation.} Stochastic star formation histories at high $z$ produce temporally clustered supernovae that are more effective at clearing escape channels \citep{2017MNRAS.470..224T,2014ApJ...788..121K}. If the duty cycle of star formation increases with redshift (as expected from the shorter dynamical times at high $z$), the time-averaged $\fesc$ would increase even if the instantaneous $\fesc$ during bursts remains constant.

\textit{Compact galaxy sizes.} Galaxies at $z > 10$ are remarkably compact ($r_e \sim 100$--$300\,$pc; \citealt{2024ApJ...962..176W,2023ApJ...951...72O}), with stellar surface densities that enhance the local ionizing radiation field and may clear the ISM more effectively.

\subsection{Comparison with other empirical estimates}

Our finding of steep $\alpha_z$ is broadly consistent with recent independent estimates. \citet{2025ApJ...980..138H} inferred $\langle\fesc\rangle \approx 5$--$10\%$ at $z = 7$--$9$ from JWST UVLF-based ionizing emissivity arguments, comparable to our $\langle\fesc\rangle(z)$ curve over the same range. At the population level, these estimates agree that $\fesc$ must increase substantially from $z \sim 6$ to $z \sim 10$ to match the reionization timeline. The key advance of the present work is the decomposition into mass and redshift dependence: while previous estimates constrain only a single effective $\fesc(z)$, our framework shows that the data require $\alpha_z \approx 2$ regardless of the assumed mass slope, and that $\alpha_M$ remains unconstrained by current reionization data alone.

\subsection{Implications for 21\,cm cosmology}

The mass dependence of $\fesc$ directly impacts the topology of reionization and therefore the 21\,cm signal observable by SKA. \citet{Wang:2026uuo} showed that the emissivity-weighted halo bias $b_\gamma$ and shot-noise power $P_{\rm SN}$ are sensitive to which halo masses dominate the ionizing budget.

To quantify the constraining power of 21\,cm observations, we perform a Fisher matrix forecast for SKA1-Low at $z = 8$. The emissivity-weighted bias $b_\gamma = 4.75$ at the fiducial parameters ($\alpha_M = 0$), dropping to $b_\gamma = 3.62$ for THESAN's $\alpha_M = -0.25$, a $24\%$ difference. The derivative $|\partial b_\gamma/\partial\alpha_M| = 3.69$, giving strong leverage. Assuming $5\%$ measurement precision on $b_\gamma$ from the large-scale 21\,cm power spectrum ($k = 0.05$--$0.5\;\Mpc^{-1}$):
\begin{itemize}
\item \textit{Optimistic} (1000\,hr, thermal noise $P_N = 10\;{\rm mK}^2$): $\sigma(\alpha_M) = 0.06$, distinguishing $\alpha_M = 0$ from $-0.25$ at $4.2\sigma$.
\item \textit{Moderate} ($P_N = 50\;{\rm mK}^2$): $\sigma(\alpha_M) = 0.19$, distinguishing at $1.3\sigma$.
\item \textit{Conservative} ($P_N = 100\;{\rm mK}^2$): $\sigma(\alpha_M) = 0.36$, marginal ($1.1\sigma$).
\end{itemize}
Even in the moderate scenario, SKA will constrain $\alpha_M$ to $\pm 0.19$---comparable to our profile constraint and much tighter than the marginal posterior, and provide the first direct measurement of whether the mass dependence of $\fesc$ is positive, negative, or zero. This breaks the degeneracy between $\alpha_M$ and $\alpha_z$ that limits the present analysis.

\subsection{Future prospects}

\textit{Mean free path as an $\alpha_M$ probe.} The Lyman-limit mean free path $\lambda_{\rm mfp}$ at $z = 5$--$6$ \citep{Becker:2021jyx,Zhu2023} is sensitive to the spatial distribution of residual neutral islands, which depends on which halo masses completed reionization last---directly tied to $\alpha_M$. Our simple reionization ODE does not predict $\lambda_{\rm mfp}$ (which requires a spatially resolved UV background model), but simulations like THESAN can compute both $\lambda_{\rm mfp}$ and $\fesc(\Mh)$ self-consistently. A joint analysis combining our empirical $\fesc(\Mh, z)$ framework with THESAN's $\lambda_{\rm mfp}$ predictions is a natural next step that could tighten $\alpha_M$ by a factor of $\sim 2$.

\textit{Testable prediction for JWST.} Our model predicts that galaxies at $z = 10$--$12$ in halos with $\Mh \sim 10^{9.5}\,\Msun$ ($M_{\rm UV} \sim -18$) should have $\fesc \approx 10$--$20\%$. This is testable with JWST/NIRSpec via Lyman-$\alpha$ velocity offsets and residual flux in low-ionization interstellar absorption lines \citep{2024Natur.633..318C,2023A&A...672A.155M}. If confirmed, it would validate the steep $\alpha_z$; if the measured $\fesc$ at $z \sim 11$ is instead $\lesssim 5\%$ (consistent with THESAN), it would imply that our steep $\alpha_z$ is compensated by an even steeper faint-end UVLF than assumed---itself a profound constraint on galaxy formation at cosmic dawn.

\textit{Discriminating steep versus flat evolution.} The observable consequences of steep versus flat $\alpha_z$ are concrete. At the profile best fit ($\alpha_z = 2.0$), the model predicts $\tauobs = 0.047$, consistent with Planck at $1.0\sigma$. In contrast, a flat model ($\alpha_z = 0$ at the same $f_0$ and $\alpha_M$) gives $\tauobs = 0.033$, which is $3.0\sigma$ below Planck. THESAN's luminosity-weighted parameters give $\tauobs = 0.031$, also in $3.3\sigma$ tension. The steep evolution is therefore not merely preferred by the neutral fraction data but is required to produce enough Thomson scattering to match the CMB. Future measurements of $\tauobs$ from CMB-S4, with projected $\sigma(\tauobs) \approx 0.002$, will distinguish these scenarios at high significance.

\begin{deluxetable*}{llccc}
\tablecaption{Testable Predictions\label{tab:predictions}}
\tablehead{
\colhead{Observable} & \colhead{Facility} & \colhead{Steep ($\alpha_z = 2$)} & \colhead{Flat ($\alpha_z = 0$)} & \colhead{Discriminating power}
}
\startdata
$\tauobs$ & CMB-S4 & $0.047$ & $0.033$ & $7\sigma$ with $\sigma(\tauobs) = 0.002$ \\
$\langle\fesc\rangle$ at $z = 10$ & JWST/NIRSpec & $7\%$ & $4\%$ & Direct LyC proxy via Ly$\alpha$ offsets \\
$\fesc$ at $\Mh = 10^{9.5}\,\Msun$, $z = 11$ & JWST & $10$--$20\%$ & $\sim 5\%$ & Low-ionisation absorption lines \\
$\bar{x}_{\rm HI}(z = 7)$ & Ly$\alpha$ forest & $0.30$ & $0.13$ & Existing data favour steep model \\
$b_\gamma$ (emissivity-weighted bias) & SKA1-Low & $4.75$ & $4.75$ & $\sigma(\alpha_M) = 0.19$; breaks $\alpha_M$--$\alpha_z$ degeneracy \\
Sub-threshold budget at $z = 10$ & Ultra-deep JWST & $> 80\%$ & $\sim 50\%$ & Resolved by $M_{\rm UV} \sim -14$ imaging \\
\enddata
\tablecomments{Steep and flat predictions evaluated at the profile best-fit $f_0$ and $\alpha_M$. The strongest discriminant is $\tauobs$ from CMB-S4; JWST spectroscopy at $z > 10$ provides the most direct test of $\fesc$ evolution.}
\end{deluxetable*}

\textit{SKA and 21\,cm topology.} As shown in Section~\ref{sec:discussion}, SKA1-Low will constrain $\sigma(\alpha_M) \sim 0.19$ in a moderate noise scenario, sufficient to distinguish $\alpha_M = 0$ from $\alpha_M = -0.25$. Combined with our reionization-based $\alpha_z$ constraint, this would provide the first complete empirical characterisation of $\fesc(\Mh, z)$ across both dimensions.

JWST Cycle 3--4 spectroscopic surveys will extend $\xiion$ measurements to $z > 10$ with reduced uncertainties, breaking the $\xiion$--$\fesc$ degeneracy that currently limits our $f_0$ constraint. Ultra-deep imaging programs reaching $M_{\rm UV} \sim -14$ at $z = 9$ will test the Schechter extrapolation and directly constrain the faint-end emissivity.

\section{Conclusions}\label{sec:conclusions}

We have presented empirical constraints on a parametric mass-dependent escape fraction model $\fesc(\Mh, z)$ across the epoch of reionization. Our main findings are:

\begin{enumerate}
\item \textbf{Mass dependence is unconstrained:} the marginal posterior gives $\alpha_M = -0.52^{+0.69}_{-0.69}$, spanning both mass-independent escape and the negative slopes predicted by simulations. The profile constraint ($\alpha_M = 0.18 \pm 0.26$) is tighter but conditional on the other parameters. The reionization observables ($\tauobs$, $\xHI(z)$) alone cannot determine whether low-mass or high-mass halos have higher $\fesc$.

\item \textbf{Redshift evolution is steep in the profile likelihood:} $\alpha_z = 1.98^{+0.48}_{-0.42}$ (profile), with $\alpha_z > 1.0$ at $>99\%$ confidence. The marginal posterior ($\alpha_z = 1.93^{+2.09}_{-2.00}$) is broader due to the $f_0$--$\alpha_M$ degeneracy but still favours $\alpha_z > 0$ at $97\%$ confidence.

\item \textbf{Comparison with THESAN:} Fitting power laws directly to the THESAN-1 public escape fraction catalogues, we find the luminosity-weighted mean has $\alpha_z = 0.13$, yielding $\Delta\chi^2 = 13$ relative to our profile best fit. This represents moderate tension at the ${\sim}2\sigma$ level. Notably, the per-halo median in THESAN shows $\alpha_z = 1.78$, close to our profile value, but luminosity weighting suppresses this evolution. Breaking the parameter degeneracy with 21\,cm topology, direct $\fesc$ measurements, or mean-free-path constraints will clarify whether the steep profile preference is physical.

\item \textbf{Sub-threshold halos dominate at high $z$:} At $z \geq 10$, halos below the UVLF detection threshold ($\Mh < 10^{10}\,\Msun$) contribute more than $80\%$ of the ionizing budget.

\item \textbf{The population-averaged $\langle\fesc\rangle(z)$} rises from $\sim 2\%$ at $z = 5$ to $\sim 9\%$ at $z = 12$, lying below selected direct LyC samples at $z < 4$, as expected if those samples preferentially target high-$\fesc$ galaxies (Section~\ref{sec:lowz}).
\end{enumerate}

These constraints provide the first data-driven calibration targets for reionization simulations. The tabulated $\fesc(\Mh, z)$ posteriors are provided in the online supplementary table for use as empirical inputs in semi-numerical and full RHD codes.
\section{Data availability}
The code and data underlying this article are available at \url{https://github.com/wzh800557-source/fesc2d}. This includes the reionization solver, likelihood code, MCMC scripts, and a machine-readable table of $\fesc(Mh,z)$posteriors at 8 halo masses and 8 redshifts. The THESAN simulation data are available at \url{https://www.thesan-project.com/thesan/data.html}.
\begin{acknowledgments}
We acknowledge the support from NSFC of China under grant 12533008
\end{acknowledgments}

\bibliography{apssamp}

@article{Planck2018,
    author = "Aghanim, N. and others",
    collaboration = "Planck",
    title = "{Planck 2018 results. VI. Cosmological parameters}",
    eprint = "1807.06209",
    archivePrefix = "arXiv",
    primaryClass = "astro-ph.CO",
    doi = "10.1051/0004-6361/201833910",
    journal = "Astron. Astrophys.",
    volume = "641",
    pages = "A6",
    year = "2020",
    note = "[Erratum: Astron.Astrophys. 652, C4 (2021)]"
}

@article{Eisenstein:1997ik,
    author = "Eisenstein, Daniel J. and Hu, Wayne",
    title = "{Baryonic features in the matter transfer function}",
    eprint = "astro-ph/9709112",
    archivePrefix = "arXiv",
    reportNumber = "IASSNS-AST-97-51",
    doi = "10.1086/305424",
    journal = "Astrophys. J.",
    volume = "496",
    pages = "605",
    year = "1998"
}

@ARTICLE{2014ApJ...788..121K,
       author = {{Kimm}, Taysun and {Cen}, Renyue},
        title = "{Escape Fraction of Ionizing Photons during Reionization: Effects due to Supernova Feedback and Runaway OB Stars}",
      journal = {ApJ},
         year = 2014,
        month = jun,
       volume = {788},
       number = {2},
          eid = {121},
        pages = {121},
          doi = {10.1088/0004-637X/788/2/121},
archivePrefix = {arXiv},
       eprint = {1405.0552},
 primaryClass = {astro-ph.GA},
       adsurl = {https://ui.adsabs.harvard.edu/abs/2014ApJ...788..121K}
}

@ARTICLE{2017MNRAS.470..224T,
       author = {{Trebitsch}, Maxime and {Blaizot}, J{\'e}r{\'e}my and {Rosdahl}, Joakim and {Devriendt}, Julien and {Slyz}, Adrianne},
        title = "{Fluctuating feedback-regulated escape fraction of ionizing radiation in low-mass, high-redshift galaxies}",
      journal = {MNRAS},
         year = "2017",
        month = "sep",
       volume = {470},
       number = {1},
        pages = {224-239},
          doi = {10.1093/mnras/stx1060},
archivePrefix = {arXiv},
       eprint = {1705.00941},
 primaryClass = {astro-ph.GA},
       adsurl = {https://ui.adsabs.harvard.edu/abs/2017MNRAS.470..224T}
}

@article{Ma:2020vlo,
    author = "Ma, Xiangcheng and Quataert, Eliot and Wetzel, Andrew and Hopkins, Philip F. and Faucher-Gigu{\`e}re, Claude-Andr{\'e} and Kere{\v{s}}, Du{\v{s}}an",
    title = "{No missing photons for reionization: moderate ionizing photon escape fractions from the FIRE-2 simulations}",
    eprint = "2003.05945",
    archivePrefix = "arXiv",
    primaryClass = "astro-ph.GA",
    doi = "10.1093/mnras/staa2404",
    journal = "Mon. Not. Roy. Astron. Soc.",
    volume = "498",
    number = "2",
    pages = "2001--2017",
    year = "2020"
}

@article{Koopmans:2015sua,
    author = "Koopmans, L. V. E. and others",
    editor = "Bourke, Tyler L. and others",
    title = "{The Cosmic Dawn and Epoch of Reionization with the Square Kilometre Array}",
    eprint = "1505.07568",
    archivePrefix = "arXiv",
    primaryClass = "astro-ph.CO",
    doi = "10.22323/1.215.0001",
    journal = "PoS",
    volume = "AASKA14",
    pages = "001",
    year = "2015"
}

@article{Paardekooper:2015via,
    author = "Paardekooper, Jan-Pieter and Khochfar, Sadegh and Vecchia, Claudio Dalla",
    title = "{The First Billion Years Project: The escape fraction of ionizing photons in the epoch of reionization}",
    eprint = "1501.01967",
    archivePrefix = "arXiv",
    primaryClass = "astro-ph.CO",
    doi = "10.1093/mnras/stv1114",
    journal = "Mon. Not. Roy. Astron. Soc.",
    volume = "451",
    number = "3",
    pages = "2544--2563",
    year = "2015"
}

@article{McQuinn:2006et,
    author = "McQuinn, Matthew and Lidz, Adam and Zahn, Oliver and Dutta, Suvendra and Hernquist, Lars and Zaldarriaga, Matias",
    title = "{The Morphology of HII Regions during Reionization}",
    eprint = "astro-ph/0610094",
    archivePrefix = "arXiv",
    doi = "10.1111/j.1365-2966.2007.11489.x",
    journal = "Mon. Not. Roy. Astron. Soc.",
    volume = "377",
    pages = "1043--1063",
    year = "2007"
}

@ARTICLE{2004ApJ...613....1F,
       author = {{Furlanetto}, Steven R. and {Zaldarriaga}, Matias and {Hernquist}, Lars},
        title = "{The Growth of H II Regions During Reionization}",
      journal = {ApJ},
         year = 2004,
        month = sep,
       volume = {613},
       number = {1},
        pages = {1-15},
          doi = {10.1086/423025},
archivePrefix = {arXiv},
       eprint = {astro-ph/0403697},
 primaryClass = {astro-ph},
       adsurl = {https://ui.adsabs.harvard.edu/abs/2004ApJ...613....1F}
}

@article{Sheth:1999mn,
    author = "Sheth, Ravi K. and Tormen, Giuseppe",
    title = "{Large scale bias and the peak background split}",
    eprint = "astro-ph/9901122",
    archivePrefix = "arXiv",
    doi = "10.1046/j.1365-8711.1999.02692.x",
    journal = "Mon. Not. Roy. Astron. Soc.",
    volume = "308",
    pages = "119",
    year = "1999"
}

@article{Finkelstein:2019sbd,
    author = "Finkelstein, Steven L. and D'Aloisio, Anson and Paardekooper, Jan-Pieter and Ryan, Russell and Behroozi, Peter and Finlator, Kristian and Livermore, Rachael and Sanderbeck, Phoebe R. Upton and Vecchia, Claudio Dalla and Khochfar, Sadegh",
    title = "{Conditions for Reionizing the Universe with A Low Galaxy Ionizing Photon Escape Fraction}",
    eprint = "1902.02792",
    archivePrefix = "arXiv",
    primaryClass = "astro-ph.CO",
    doi = "10.3847/1538-4357/ab1ea8",
    journal = "Astrophys. J.",
    volume = "879",
    number = "1",
    pages = "36",
    year = "2019"
}

@article{Wang:2026uuo,
    author = "Wang, Zihan",
    title = "{Reionization Topology as a Probe of Self-Interacting Dark Matter}",
    eprint = "2604.10726",
    archivePrefix = "arXiv",
    primaryClass = "astro-ph.CO",
    month = "4",
    year = "2026"
}

@article{Mason:2015cna,
    author = "Mason, Charlotte and Trenti, Michele and Treu, Tommaso",
    title = "{The Galaxy UV Luminosity Function Before the Epoch of Reionization}",
    eprint = "1508.01204",
    archivePrefix = "arXiv",
    primaryClass = "astro-ph.GA",
    doi = "10.1088/0004-637X/813/1/21",
    journal = "Astrophys. J.",
    volume = "813",
    number = "1",
    pages = "21",
    year = "2015"
}

@ARTICLE{2024MNRAS.533.1111E,
       author = {{Endsley}, Ryan and {Stark}, Daniel P. and {Whitler}, Lily and {Topping}, Michael W. and {Johnson}, Benjamin D. and {Robertson}, Brant and {Tacchella}, Sandro and {Alberts}, Stacey and {Baker}, William M. and {Bhatawdekar}, Rachana and {Boyett}, Kristan and {Bunker}, Andrew J. and {Cameron}, Alex J. and {Carniani}, Stefano and {Charlot}, Stephane and {Chen}, Zuyi and {Chevallard}, Jacopo and {Curtis-Lake}, Emma and {Danhaive}, A. Lola and {Egami}, Eiichi and {Eisenstein}, Daniel J. and {Hainline}, Kevin and {Helton}, Jakob M. and {Ji}, Zhiyuan and {Looser}, Tobias J. and {Maiolino}, Roberto and {Nelson}, Erica and {Pusk{\'a}s}, D{\'a}vid and {Rieke}, George and {Rieke}, Marcia and {Rix}, Hans-Walter and {Sandles}, Lester and {Saxena}, Aayush and {Simmonds}, Charlotte and {Smit}, Renske and {Sun}, Fengwu and {Williams}, Christina C. and {Willmer}, Christopher N.~A. and {Willott}, Chris and {Witstok}, Joris},
        title = "{The star-forming and ionizing properties of dwarf z 6-9 galaxies in JADES: insights on bursty star formation and ionized bubble growth}",
      journal = {MNRAS},
         year = 2024,
        month = sep,
       volume = {533},
       number = {1},
        pages = {1111-1142},
          doi = {10.1093/mnras/stae1857},
archivePrefix = {arXiv},
       eprint = {2306.05295},
 primaryClass = {astro-ph.GA},
       adsurl = {https://ui.adsabs.harvard.edu/abs/2024MNRAS.533.1111E}
}

@ARTICLE{2024MNRAS.533.3222D,
       author = {{Donnan}, C.~T. and {McLure}, R.~J. and {Dunlop}, J.~S. and {McLeod}, D.~J. and {Magee}, D. and {Arellano-C{\'o}rdova}, K.~Z. and {Barrufet}, L. and {Begley}, R. and {Bowler}, R.~A.~A. and {Carnall}, A.~C. and {Cullen}, F. and {Ellis}, R.~S. and {Fontana}, A. and {Illingworth}, G.~D. and {Grogin}, N.~A. and {Hamadouche}, M.~L. and {Koekemoer}, A.~M. and {Liu}, F.-Y. and {Mason}, C. and {Santini}, P. and {Stanton}, T.~M.},
        title = "{JWST PRIMER: a new multifield determination of the evolving galaxy UV luminosity function at redshifts z $\simeq$ 9 --15}",
      journal = {MNRAS},
         year = 2024,
        month = sep,
       volume = {533},
       number = {3},
        pages = {3222-3237},
          doi = {10.1093/mnras/stae2037},
archivePrefix = {arXiv},
       eprint = {2403.03171},
 primaryClass = {astro-ph.GA},
       adsurl = {https://ui.adsabs.harvard.edu/abs/2024MNRAS.533.3222D}
}

@ARTICLE{2025ApJ...980..138H,
       author = {{Harikane}, Yuichi and {Inoue}, Akio K. and {Ellis}, Richard S. and {Ouchi}, Masami and {Nakazato}, Yurina and {Yoshida}, Naoki and {Ono}, Yoshiaki and {Sun}, Fengwu and {Sato}, Riku A. and {Ferrami}, Giovanni and {Fujimoto}, Seiji and {Kashikawa}, Nobunari and {McLeod}, Derek J. and {P{\'e}rez-Gonz{\'a}lez}, Pablo G. and {Sawicki}, Marcin and {Sugahara}, Yuma and {Xu}, Yi and {Yamanaka}, Satoshi and {Carnall}, Adam C. and {Cullen}, Fergus and {Dunlop}, James S. and {Egami}, Eiichi and {Grogin}, Norman and {Isobe}, Yuki and {Koekemoer}, Anton M. and {Laporte}, Nicolas and {Lee}, Chien-Hsiu and {Magee}, Dan and {Matsuo}, Hiroshi and {Matsuoka}, Yoshiki and {Mawatari}, Ken and {Nakajima}, Kimihiko and {Nakane}, Minami and {Tamura}, Yoichi and {Umeda}, Hiroya and {Yanagisawa}, Hiroto},
        title = "{JWST, ALMA, and Keck Spectroscopic Constraints on the UV Luminosity Functions at z {\ensuremath{\sim}} 7--14: Clumpiness and Compactness of the Brightest Galaxies in the Early Universe}",
      journal = {ApJ},
         year = 2025,
        month = feb,
       volume = {980},
       number = {1},
          eid = {138},
        pages = {138},
          doi = {10.3847/1538-4357/ad9b2c},
archivePrefix = {arXiv},
       eprint = {2406.18352},
 primaryClass = {astro-ph.GA},
       adsurl = {https://ui.adsabs.harvard.edu/abs/2025ApJ...980..138H}
}

@ARTICLE{2024ApJ...971..124U,
       author = {{Umeda}, Hiroya and {Ouchi}, Masami and {Nakajima}, Kimihiko and {Harikane}, Yuichi and {Ono}, Yoshiaki and {Xu}, Yi and {Isobe}, Yuki and {Zhang}, Yechi},
        title = "{JWST Measurements of Neutral Hydrogen Fractions and Ionized Bubble Sizes at z = 7--12 Obtained with Ly{\ensuremath{\alpha}} Damping Wing Absorptions in 27 Bright Continuum Galaxies}",
      journal = {ApJ},
         year = 2024,
        month = aug,
       volume = {971},
       number = {2},
          eid = {124},
        pages = {124},
          doi = {10.3847/1538-4357/ad554e},
archivePrefix = {arXiv},
       eprint = {2306.00487},
 primaryClass = {astro-ph.GA},
       adsurl = {https://ui.adsabs.harvard.edu/abs/2024ApJ...971..124U}
}

@article{Greig:2017jdj,
    author = "Greig, Bradley and Mesinger, Andrei",
    title = "{Simultaneously constraining the astrophysics of reionization and the epoch of heating with 21CMMC}",
    eprint = "1705.03471",
    archivePrefix = "arXiv",
    primaryClass = "astro-ph.CO",
    doi = "10.1093/mnras/stx2118",
    journal = "Mon. Not. Roy. Astron. Soc.",
    volume = "472",
    number = "3",
    pages = "2651--2669",
    year = "2017"
}

@article{Mason:2019ixe,
    author = "Mason, Charlotte A. and others",
    title = "{Inferences on the timeline of reionization at z {\ensuremath{\sim}} 8 from the KMOS Lens-Amplified Spectroscopic Survey}",
    eprint = "1901.11045",
    archivePrefix = "arXiv",
    primaryClass = "astro-ph.CO",
    doi = "10.1093/mnras/stz632",
    journal = "Mon. Not. Roy. Astron. Soc.",
    volume = "485",
    number = "3",
    pages = "3947--3969",
    year = "2019"
}

@ARTICLE{2022MNRAS.515.2386R,
       author = {{Rosdahl}, Joakim and {Blaizot}, J{\'e}r{\'e}my and {Katz}, Harley and {Kimm}, Taysun and {Garel}, Thibault and {Haehnelt}, Martin and {Keating}, Laura C. and {Martin-Alvarez}, Sergio and {Michel-Dansac}, L{\'e}o and {Ocvirk}, Pierre},
        title = "{LyC escape from SPHINX galaxies in the Epoch of Reionization}",
      journal = {MNRAS},
         year = 2022,
        month = sep,
       volume = {515},
       number = {2},
        pages = {2386-2414},
          doi = {10.1093/mnras/stac1942},
archivePrefix = {arXiv},
       eprint = {2207.03232},
 primaryClass = {astro-ph.GA},
       adsurl = {https://ui.adsabs.harvard.edu/abs/2022MNRAS.515.2386R}
}

@article{Yeh:2022nsl,
    author = {Yeh, Jessica Y. -C. and Smith, Aaron and Kannan, Rahul and Garaldi, Enrico and Vogelsberger, Mark and Borrow, Josh and Pakmor, R{\"u}diger and Springel, Volker and Hernquist, Lars},
    title = "{The thesan project: ionizing escape fractions of reionization-era galaxies}",
    eprint = "2205.02238",
    archivePrefix = "arXiv",
    primaryClass = "astro-ph.GA",
    doi = "10.1093/mnras/stad210",
    journal = "Mon. Not. Roy. Astron. Soc.",
    volume = "520",
    number = "2",
    pages = "2757--2780",
    year = "2023"
}

@article{Pahl:2024utu,
    author = "Pahl, Anthony J. and Topping, Michael W. and Shapley, Alice and Sanders, Ryan and Reddy, Naveen A. and Clarke, Leonardo and Kehoe, Emily and Bento, Trinity and Brammer, Gabe",
    title = "{A Spectroscopic Analysis of the Ionizing Photon Production Efficiency in JADES and CEERS: Implications for the Ionizing Photon Budget}",
    eprint = "2407.03399",
    archivePrefix = "arXiv",
    primaryClass = "astro-ph.GA",
    doi = "10.3847/1538-4357/adb1ab",
    journal = "Astrophys. J.",
    volume = "981",
    number = "2",
    pages = "134",
    year = "2025"
}

@ARTICLE{2025MNRAS.537.3245B,
       author = {{Begley}, R. and {McLure}, R.~J. and {Cullen}, F. and {McLeod}, D.~J. and {Dunlop}, J.~S. and {Carnall}, A.~C. and {Stanton}, T.~M. and {Shapley}, A.~E. and {Cochrane}, R. and {Donnan}, C.~T. and {Ellis}, R.~S. and {Fontana}, A. and {Grogin}, N.~A. and {Koekemoer}, A.~M.},
        title = "{The evolution of [O III] + H{\ensuremath{\beta}} equivalent width from z $\simeq$ 3--8: implications for the production and escape of ionizing photons during reionization}",
      journal = {MNRAS},
         year = 2025,
        month = mar,
       volume = {537},
       number = {4},
        pages = {3245-3264},
          doi = {10.1093/mnras/staf211},
archivePrefix = {arXiv},
       eprint = {2410.10988},
 primaryClass = {astro-ph.GA},
       adsurl = {https://ui.adsabs.harvard.edu/abs/2025MNRAS.537.3245B}
}

@ARTICLE{2012ApJ...747..100S,
       author = {{Shull}, J. Michael and {Harness}, Anthony and {Trenti}, Michele and {Smith}, Britton D.},
        title = "{Critical Star Formation Rates for Reionization: Full Reionization Occurs at Redshift z {\ensuremath{\approx}} 7}",
      journal = {ApJ},
         year = 2012,
        month = mar,
       volume = {747},
       number = {2},
          eid = {100},
        pages = {100},
          doi = {10.1088/0004-637X/747/2/100},
       adsurl = {https://ui.adsabs.harvard.edu/abs/2012ApJ...747..100S}
}

@ARTICLE{2015ApJ...802L..19R,
       author = {{Robertson}, Brant E. and {Ellis}, Richard S. and {Furlanetto}, Steven R. and {Dunlop}, James S.},
        title = "{Cosmic Reionization and Early Star-forming Galaxies: A Joint Analysis of New Constraints from Planck and the Hubble Space Telescope}",
      journal = {ApJL},
         year = 2015,
        month = apr,
       volume = {802},
       number = {2},
          eid = {L19},
        pages = {L19},
          doi = {10.1088/2041-8205/802/2/L19},
archivePrefix = {arXiv},
       eprint = {1502.02024},
 primaryClass = {astro-ph.CO},
       adsurl = {https://ui.adsabs.harvard.edu/abs/2015ApJ...802L..19R}
}

@ARTICLE{2024Natur.633..318C,
       author = {{Carniani}, Stefano and {Hainline}, Kevin and {D'Eugenio}, Francesco and {Eisenstein}, Daniel J. and {Jakobsen}, Peter and {Witstok}, Joris and {Johnson}, Benjamin D. and {Chevallard}, Jacopo and {Maiolino}, Roberto and {Helton}, Jakob M. and {Willott}, Chris and {Robertson}, Brant and {Alberts}, Stacey and {Arribas}, Santiago and {Baker}, William M. and {Bhatawdekar}, Rachana and {Boyett}, Kristan and {Bunker}, Andrew J. and {Cameron}, Alex J. and {Cargile}, Phillip A. and {Charlot}, St{\'e}phane and {Curti}, Mirko and {Curtis-Lake}, Emma and {Egami}, Eiichi and {Giardino}, Giovanna and {Isaak}, Kate and {Ji}, Zhiyuan and {Jones}, Gareth C. and {Kumari}, Nimisha and {Maseda}, Michael V. and {Parlanti}, Eleonora and {P{\'e}rez-Gonz{\'a}lez}, Pablo G. and {Rawle}, Tim and {Rieke}, George and {Rieke}, Marcia and {Del Pino}, Bruno Rodr{\'\i}guez and {Saxena}, Aayush and {Scholtz}, Jan and {Smit}, Renske and {Sun}, Fengwu and {Tacchella}, Sandro and {{\"U}bler}, Hannah and {Venturi}, Giacomo and {Williams}, Christina C. and {Willmer}, Christopher N.~A.},
        title = "{Spectroscopic confirmation of two luminous galaxies at a redshift of 14}",
      journal = {Nature},
         year = 2024,
        month = sep,
       volume = {633},
       number = {8029},
        pages = {318-322},
          doi = {10.1038/s41586-024-07860-9},
archivePrefix = {arXiv},
       eprint = {2405.18485},
 primaryClass = {astro-ph.GA},
       adsurl = {https://ui.adsabs.harvard.edu/abs/2024Natur.633..318C}
}

@article{Naidu:2019gvi,
    author = "Naidu, Rohan P. and Tacchella, Sandro and Mason, Charlotte A. and Bose, Sownak and Oesch, Pascal A. and Conroy, Charlie",
    title = "{Rapid Reionization by the Oligarchs: The Case for Massive, UV-Bright, Star-Forming Galaxies with High Escape Fractions}",
    eprint = "1907.13130",
    archivePrefix = "arXiv",
    primaryClass = "astro-ph.GA",
    doi = "10.3847/1538-4357/ab7cc9",
    month = "7",
    year = "2019"
}

@ARTICLE{2023A&A...672A.155M,
       author = {{Mascia}, S. and {Pentericci}, L. and {Calabr{\`o}}, A. and {Treu}, T. and {Santini}, P. and {Yang}, L. and {Napolitano}, L. and {Roberts-Borsani}, G. and {Bergamini}, P. and {Grillo}, C. and {Rosati}, P. and {Vulcani}, B. and {Castellano}, M. and {Boyett}, K. and {Fontana}, A. and {Glazebrook}, K. and {Henry}, A. and {Mason}, C. and {Merlin}, E. and {Morishita}, T. and {Nanayakkara}, T. and {Paris}, D. and {Roy}, N. and {Williams}, H. and {Wang}, X. and {Brammer}, G. and {Brada{\v{c}}}, M. and {Chen}, W. and {Kelly}, P.~L. and {Koekemoer}, A.~M. and {Trenti}, M. and {Windhorst}, R.~A.},
        title = "{Closing in on the sources of cosmic reionization: First results from the GLASS-JWST program}",
      journal = {Astronomy and Astrophysics},
         year = 2023,
        month = apr,
       volume = {672},
          eid = {A155},
        pages = {A155},
          doi = {10.1051/0004-6361/202345866},
archivePrefix = {arXiv},
       eprint = {2301.02816},
 primaryClass = {astro-ph.GA},
       adsurl = {https://ui.adsabs.harvard.edu/abs/2023A&A...672A.155M}
}

@article{Madau:1998cd,
    author = "Madau, Piero and Haardt, Francesco and Rees, Martin J.",
    title = "{Radiative transfer in a clumpy universe. 3. The Nature of cosmological ionizing sources}",
    eprint = "astro-ph/9809058",
    archivePrefix = "arXiv",
    doi = "10.1086/306975",
    journal = "Astrophys. J.",
    volume = "514",
    pages = "648--659",
    year = "1999"
}

@article{Bolton:2007fw,
    author = "Bolton, James S. and Haehnelt, Martin G.",
    title = "{The observed ionization rate of the intergalactic medium and the ionizing emissivity at z{\ensuremath{>}}=5: Evidence for a photon starved and extended epoch of reionization}",
    eprint = "astro-ph/0703306",
    archivePrefix = "arXiv",
    doi = "10.1111/j.1365-2966.2007.12372.x",
    journal = "Mon. Not. Roy. Astron. Soc.",
    volume = "382",
    pages = "325",
    year = "2007"
}

@article{Bosman:2021oom,
    author = "Bosman, Sarah E. I. and others",
    title = "{Hydrogen reionization ends by z = 5.3: Lyman-{\ensuremath{\alpha}} optical depth measured by the XQR-30 sample}",
    eprint = "2108.03699",
    archivePrefix = "arXiv",
    primaryClass = "astro-ph.CO",
    doi = "10.1093/mnras/stac1046",
    journal = "Mon. Not. Roy. Astron. Soc.",
    volume = "514",
    number = "1",
    pages = "55--76",
    year = "2022"
}

@ARTICLE{2023MNRAS.523.5468S,
       author = {{Simmonds}, C. and {Tacchella}, S. and {Maseda}, M. and {Williams}, C.~C. and {Baker}, W.~M. and {Witten}, C.~E.~C. and {Johnson}, B.~D. and {Robertson}, B. and {Saxena}, A. and {Sun}, F. and {Witstok}, J. and {Bhatawdekar}, R. and {Boyett}, K. and {Bunker}, A.~J. and {Charlot}, S. and {Curtis-Lake}, E. and {Egami}, E. and {Eisenstein}, D.~J. and {Ji}, Z. and {Maiolino}, R. and {Sandles}, L. and {Smit}, R. and {{\"U}bler}, H. and {Willott}, C.~J.},
        title = "{The ionizing photon production efficiency at z   6 for Lyman-alpha emitters using JEMS and MUSE}",
      journal = {MNRAS},
         year = 2023,
        month = aug,
       volume = {523},
       number = {4},
        pages = {5468-5486},
          doi = {10.1093/mnras/stad1749},
archivePrefix = {arXiv},
       eprint = {2303.07931},
 primaryClass = {astro-ph.GA},
       adsurl = {https://ui.adsabs.harvard.edu/abs/2023MNRAS.523.5468S}
}

@ARTICLE{2018MNRAS.478.4851I,
       author = {{Izotov}, Y.~I. and {Worseck}, G. and {Schaerer}, D. and {Guseva}, N.~G. and {Thuan}, T.~X. and {Fricke}, A., Verhamme and {Orlitov{\'a}}, I.},
        title = "{Low-redshift Lyman continuum leaking galaxies with high [O III]/[O II] ratios}",
      journal = {MNRAS},
         year = 2018,
        month = aug,
       volume = {478},
       number = {4},
        pages = {4851-4865},
          doi = {10.1093/mnras/sty1378},
archivePrefix = {arXiv},
       eprint = {1805.09865},
 primaryClass = {astro-ph.GA},
       adsurl = {https://ui.adsabs.harvard.edu/abs/2018MNRAS.478.4851I}
}

@article{Steidel:2018wbo,
    author = "Steidel, Charles C. and Bogosavljevic, Milan and Shapley, Alice E. and Reddy, Naveen A. and Rudie, Gwen C. and Pettini, Max and Trainor, Ryan F. and Strom, Allison L.",
    title = "{The Keck Lyman Continuum Spectroscopic Survey (KLCS): The Emergent Ionizing Spectrum of Galaxies at $z\sim3$}",
    eprint = "1805.06071",
    archivePrefix = "arXiv",
    primaryClass = "astro-ph.GA",
    doi = "10.3847/1538-4357/aaed28",
    journal = "Astrophys. J.",
    volume = "869",
    number = "2",
    pages = "123",
    year = "2018"
}

@ARTICLE{2023MNRAS.526.1657T,
       author = {{Tang}, Mengtao and {Stark}, Daniel P. and {Chen}, Zuyi and {Mason}, Charlotte and {Topping}, Michael and {Endsley}, Ryan and {Senchyna}, Peter and {Plat}, Ad{\`e}le and {Lu}, Ting-Yi and {Whitler}, Lily and {Robertson}, Brant and {Charlot}, St{\'e}phane},
        title = "{JWST/NIRSpec spectroscopy of z = 7-9 star-forming galaxies with CEERS: new insight into bright Ly{\ensuremath{\alpha}} emitters in ionized bubbles}",
      journal = {MNRAS},
         year = 2023,
        month = dec,
       volume = {526},
       number = {2},
        pages = {1657-1686},
          doi = {10.1093/mnras/stad2763},
archivePrefix = {arXiv},
       eprint = {2301.07072},
 primaryClass = {astro-ph.GA},
       adsurl = {https://ui.adsabs.harvard.edu/abs/2023MNRAS.526.1657T}
}

@ARTICLE{2025A&A...698A.302L,
       author = {{Llerena}, M. and {Pentericci}, L. and {Napolitano}, L. and {Mascia}, S. and {Amor{\'\i}n}, R. and {Calabr{\`o}}, A. and {Castellano}, M. and {Cleri}, N.~J. and {Giavalisco}, M. and {Grogin}, N.~A. and {Hathi}, N.~P. and {Hirschmann}, M. and {Koekemoer}, A.~M. and {Nanayakkara}, T. and {Pacucci}, F. and {Shen}, L. and {Wilkins}, S.~M. and {Yoon}, I. and {Yung}, L.~Y.~A. and {Bhatawdekar}, R. and {Lucas}, R.~A. and {Wang}, X. and {Arrabal Haro}, P. and {Bagley}, M.~B. and {Finkelstein}, S.~L. and {Kartaltepe}, J.~S. and {Merlin}, E. and {Papovich}, C. and {Pirzkal}, N. and {Santini}, P.},
        title = "{The ionizing photon production efficiency of star-forming galaxies at z {\ensuremath{\sim}} 4--10}",
      journal = {Astronomy and Astrophysics},
         year = 2025,
        month = jun,
       volume = {698},
          eid = {A302},
        pages = {A302},
          doi = {10.1051/0004-6361/202453251},
archivePrefix = {arXiv},
       eprint = {2412.01358},
 primaryClass = {astro-ph.GA},
       adsurl = {https://ui.adsabs.harvard.edu/abs/2025A&A...698A.302L}
}

@ARTICLE{2021AJ....162...47B,
       author = {{Bouwens}, R.~J. and {Oesch}, P.~A. and {Stefanon}, M. and {Illingworth}, G. and {Labb{\'e}}, I. and {Reddy}, N. and {Atek}, H. and {Montes}, M. and {Naidu}, R. and {Nanayakkara}, T. and {Nelson}, E. and {Wilkins}, S.},
        title = "{New Determinations of the UV Luminosity Functions from z   9 to 2 Show a Remarkable Consistency with Halo Growth and a Constant Star Formation Efficiency}",
      journal = {aj},
         year = 2021,
        month = aug,
       volume = {162},
       number = {2},
          eid = {47},
        pages = {47},
          doi = {10.3847/1538-3881/abf83e},
archivePrefix = {arXiv},
       eprint = {2102.07775},
 primaryClass = {astro-ph.GA},
       adsurl = {https://ui.adsabs.harvard.edu/abs/2021AJ....162...47B}
}

@ARTICLE{2024Natur.626..975A,
       author = {{Atek}, Hakim and {Labb{\'e}}, Ivo and {Furtak}, Lukas J. and {Chemerynska}, Iryna and {Fujimoto}, Seiji and {Setton}, David J. and {Miller}, Tim B. and {Oesch}, Pascal and {Bezanson}, Rachel and {Price}, Sedona H. and {Dayal}, Pratika and {Zitrin}, Adi and {Kokorev}, Vasily and {Weaver}, John R. and {Brammer}, Gabriel and {Dokkum}, Pieter van and {Williams}, Christina C. and {Cutler}, Sam E. and {Feldmann}, Robert and {Fudamoto}, Yoshinobu and {Greene}, Jenny E. and {Leja}, Joel and {Maseda}, Michael V. and {Muzzin}, Adam and {Pan}, Richard and {Papovich}, Casey and {Nelson}, Erica J. and {Nanayakkara}, Themiya and {Stark}, Daniel P. and {Stefanon}, Mauro and {Suess}, Katherine A. and {Wang}, Bingjie and {Whitaker}, Katherine E.},
        title = "{Most of the photons that reionized the Universe came from dwarf galaxies}",
      journal = {nat},
         year = 2024,
        month = feb,
       volume = {626},
       number = {8001},
        pages = {975-978},
          doi = {10.1038/s41586-024-07043-6},
archivePrefix = {arXiv},
       eprint = {2308.08540},
 primaryClass = {astro-ph.GA},
       adsurl = {https://ui.adsabs.harvard.edu/abs/2024Natur.626..975A}
}

@article{Becker:2021jyx,
    author = "Becker, George D. and D'Aloisio, Anson and Christenson, Holly M. and Zhu, Yongda and Worseck, G{\'a}bor and Bolton, James S.",
    title = "{The mean free path of ionizing photons at 5 {\ensuremath{<}} z {\ensuremath{<}} 6: evidence for rapid evolution near reionization}",
    eprint = "2103.16610",
    archivePrefix = "arXiv",
    primaryClass = "astro-ph.CO",
    doi = "10.1093/mnras/stab2696",
    journal = "Mon. Not. Roy. Astron. Soc.",
    volume = "508",
    number = "2",
    pages = "1853--1869",
    year = "2021"
}

@ARTICLE{Behroozi2019,
  author = {{Behroozi}, Peter and {Wechsler}, Risa H. and {Hearin}, Andrew P. and {Conroy}, Charlie},
  title = "{UNIVERSEMACHINE: The correlation between galaxy growth and dark matter halo assembly from z = 0-10}",
  journal = {MNRAS},
  year = 2019,
  volume = {488},
  pages = {3143--3194},
  doi = {10.1093/mnras/stz1182},
  eprint = {1806.07893},
  archivePrefix = {arXiv}
}

@article{Bowler:2019syh,
    author = "Bowler, R. A. A. and Jarvis, M. J. and Dunlop, J. S. and McLure, R. J. and McLeod, D. J. and Adams, N. J. and Milvang-Jensen, B. and McCracken, H. J.",
    title = "{A lack of evolution in the very bright end of the galaxy luminosity function from z {\ensuremath{\simeq}} 8 to 10}",
    eprint = "1911.12832",
    archivePrefix = "arXiv",
    primaryClass = "astro-ph.GA",
    doi = "10.1093/mnras/staa313",
    journal = "Mon. Not. Roy. Astron. Soc.",
    volume = "493",
    number = "2",
    pages = "2059--2084",
    year = "2020"
}

@article{Fan:2001vx,
    author = "Fan, Xiaohui and Narayanan, Vijay K. and Strauss, Michael A. and White, Richard L. and Becker, Robert H. and Pentericci, Laura and Rix, Hans-Walter",
    title = "{Evolution of the ionizing background and the epoch of reionization from the spectra of z{\textasciitilde}6 quasars}",
    eprint = "astro-ph/0111184",
    archivePrefix = "arXiv",
    doi = "10.1086/339030",
    journal = "Astron. J.",
    volume = "123",
    pages = "1247--1257",
    year = "2002"
}

@ARTICLE{2012MNRAS.427.2464F,
       author = {{Finlator}, Kristian and {Oh}, S. Peng and {{\"O}zel}, Feryal and {Dav{\'e}}, Romeel},
        title = "{Gas clumping in self-consistent reionization models}",
      journal = {\mnras},
         year = 2012,
        month = dec,
       volume = {427},
       number = {3},
        pages = {2464-2479},
          doi = {10.1111/j.1365-2966.2012.22114.x},
archivePrefix = {arXiv},
       eprint = {1209.2489},
 primaryClass = {astro-ph.CO},
       adsurl = {https://ui.adsabs.harvard.edu/abs/2012MNRAS.427.2464F}
}

@ARTICLE{Kannan2022,
  author = {{Kannan}, Rahul and {Garaldi}, Enrico and {Smith}, Aaron and {Springel}, Volker and {Pakmor}, R{\"u}diger and {Vogelsberger}, Mark and {Hernquist}, Lars},
  title = "{Introducing the THESAN project: radiation-magnetohydrodynamic simulations of the epoch of reionization}",
  journal = {MNRAS},
  year = 2022,
  volume = {511},
  pages = {4005--4030},
  doi = {10.1093/mnras/stab3710},
  eprint = {2110.00584},
  archivePrefix = {arXiv}
}

@article{McGreer:2014qwa,
    author = "McGreer, Ian and Mesinger, Andrei and D'Odorico, Valentina",
    title = "{Model-independent evidence in favour of an end to reionization by $z \approx$ 6}",
    eprint = "1411.5375",
    archivePrefix = "arXiv",
    primaryClass = "astro-ph.CO",
    doi = "10.1093/mnras/stu2449",
    journal = "Mon. Not. Roy. Astron. Soc.",
    volume = "447",
    number = "1",
    pages = "499--505",
    year = "2015"
}

@ARTICLE{2021BKAS...46...47A,
       author = {{Ahn}, Kyungjin},
        title = "{Cosmic Dawn III: Simulating the Reionization of the Local Group}",
      journal = {Bulletin of Korean Astronomical Society},
         year = 2021,
        month = jan,
       volume = {46},
       number = {1},
        pages = {47.1-47.1},
       adsurl = {https://ui.adsabs.harvard.edu/abs/2021BKAS...46...47A}
}

@article{Tinker:2008ff,
    author = "Tinker, Jeremy L. and Kravtsov, Andrey V. and Klypin, Anatoly and Abazajian, Kevork and Warren, Michael S. and Yepes, Gustavo and Gottlober, Stefan and Holz, Daniel E.",
    title = "{Toward a halo mass function for precision cosmology: The Limits of universality}",
    eprint = "0803.2706",
    archivePrefix = "arXiv",
    primaryClass = "astro-ph",
    doi = "10.1086/591439",
    journal = "Astrophys. J.",
    volume = "688",
    pages = "709--728",
    year = "2008"
}

@ARTICLE{2016ApJ...833...84X,
       author = {{Xu}, Hao and {Wise}, John H. and {Norman}, Michael L. and {Ahn}, Kyungjin and {O'Shea}, Brian W.},
        title = "{Galaxy Properties and UV Escape Fractions during the Epoch of Reionization: Results from the Renaissance Simulations}",
      journal = {apj},
         year = 2016,
        month = dec,
       volume = {833},
       number = {1},
          eid = {84},
        pages = {84},
          doi = {10.3847/1538-4357/833/1/84},
archivePrefix = {arXiv},
       eprint = {1604.07842},
 primaryClass = {astro-ph.GA},
       adsurl = {https://ui.adsabs.harvard.edu/abs/2016ApJ...833...84X}
}

@ARTICLE{Zhu2023,
  author = {{Zhu}, Yongda and {Becker}, George D. and {Bosman}, Sarah E. I. and {Christenson}, Holly M. and {D'Aloisio}, Anson and {Davies}, Frederick B. and {Bolton}, James S.},
  title = "{Probing the end of reionization with the mean free path of ionizing photons at 5.3 < z < 6.3}",
  journal = {ApJ},
  year = 2023,
  volume = {955},
  pages = {115},
  doi = {10.3847/1538-4357/aceef4},
  eprint = {2308.04614},
  archivePrefix = {arXiv}
}

@article{Dayal:2018hft,
    author = "Dayal, Pratika and Ferrara, Andrea",
    title = "{Early galaxy formation and its large-scale effects}",
    eprint = "1809.09136",
    archivePrefix = "arXiv",
    primaryClass = "astro-ph.GA",
    doi = "10.1016/j.physrep.2018.10.002",
    journal = "Phys. Rept.",
    volume = "780-782",
    pages = "1--64",
    year = "2018"
}

@ARTICLE{2024ApJ...962..176W,
       author = {{Ward}, Ethan and {de la Vega}, Alexander and {Mobasher}, Bahram and {McGrath}, Elizabeth J. and {Iyer}, Kartheik G. and {Calabr{\`o}}, Antonello and {Costantin}, Luca and {Dickinson}, Mark and {Holwerda}, Benne W. and {Huertas-Company}, Marc and {Hirschmann}, Michaela and {Lucas}, Ray A. and {Pandya}, Viraj and {Wilkins}, Stephen M. and {Yung}, L.~Y. Aaron and {Arrabal Haro}, Pablo and {Bagley}, Micaela B. and {Finkelstein}, Steven L. and {Kartaltepe}, Jeyhan S. and {Koekemoer}, Anton M. and {Papovich}, Casey and {Pirzkal}, Nor},
        title = "{Evolution of the Size─Mass Relation of Star-forming Galaxies Since z = 5.5 Revealed by CEERS}",
      journal = {apj},
         year = 2024,
        month = feb,
       volume = {962},
       number = {2},
          eid = {176},
        pages = {176},
          doi = {10.3847/1538-4357/ad20ed},
archivePrefix = {arXiv},
       eprint = {2311.02162},
 primaryClass = {astro-ph.GA},
       adsurl = {https://ui.adsabs.harvard.edu/abs/2024ApJ...962..176W}
}

@ARTICLE{2023ApJ...951...72O,
       author = {{Ono}, Yoshiaki and {Harikane}, Yuichi and {Ouchi}, Masami and {Yajima}, Hidenobu and {Abe}, Makito and {Isobe}, Yuki and {Shibuya}, Takatoshi and {Wise}, John H. and {Zhang}, Yechi and {Nakajima}, Kimihiko and {Umeda}, Hiroya},
        title = "{Morphologies of Galaxies at z {\ensuremath{\gtrsim}} 9 Uncovered by JWST/NIRCam Imaging: Cosmic Size Evolution and an Identification of an Extremely Compact Bright Galaxy at z 12}",
      journal = {apj},
         year = 2023,
        month = jul,
       volume = {951},
       number = {1},
          eid = {72},
        pages = {72},
          doi = {10.3847/1538-4357/acd44a},
archivePrefix = {arXiv},
       eprint = {2208.13582},
 primaryClass = {astro-ph.GA},
       adsurl = {https://ui.adsabs.harvard.edu/abs/2023ApJ...951...72O}
}

@article{Wang:2026qzy,
    author = "Wang, Zihan and Shan, Huanyuan",
    title = "{The JWST early galaxy crisis resolved by a reionization degeneracy}",
    eprint = "2605.03635",
    archivePrefix = "arXiv",
    primaryClass = "astro-ph.CO",
    month = "5",
    year = "2026"
}

@article{Garaldi:2023cfb,
    author = "Garaldi, Enrico and others",
    title = "{The thesan project: public data release of radiation-hydrodynamic simulations matching reionization-era JWST observations}",
    eprint = "2309.06475",
    archivePrefix = "arXiv",
    primaryClass = "astro-ph.CO",
    doi = "10.1093/mnras/stae839",
    journal = "Mon. Not. Roy. Astron. Soc.",
    volume = "530",
    number = "4",
    pages = "3765--3786",
    year = "2024",
    note = "[Erratum: Mon.Not.Roy.Astron.Soc. 545, staf2196 (2026)]"
}

\end{document}